\documentclass[letterpaper,12pt,notoc]{JHEP3}
\def\pd{\partial}
\def\mc{\mathcal}

\usepackage{graphicx}
\usepackage{amsmath}
\usepackage{amssymb}

\preprint{ \hbox{}\hfill arXiv:1311.7581}

\title{Deformations of large $N=(4,4)$ 2D SCFT from 3D
gauged supergravity}
\author{Parinya Karndumri\\
String Theory and Supergravity Group, Department
of Physics, Faculty of Science, Chulalongkorn University, 254 Phayathai Road, Pathumwan, Bangkok 10330, Thailand\\
Thailand Center of Excellence in Physics, CHE, Ministry of
Education, Bangkok 10400, Thailand
\\
E-mail: \email{parinya.ka@hotmail.com}}

\abstract{Supersymmetric and non-supersymmetric deformations of
large $N=(4,4)$ SCFT with superconformal symmetry
$D^1(2,1;\alpha)\times D^1(2,1;\alpha)$ are explored in the gravity
dual described by a Chern-Simons $N=8$, $(SO(4)\times SO(4))\ltimes
\mathbf{T}^{12}$ gauged supergravity in three dimensions. For
$\alpha>0$, the gauged supergravity describes an effective theory of
the maximal supergravity in nine dimensions on $AdS_3\times
S^3\times S^3$ with the parameter $\alpha$ being the ratio of the
two $S^3$ radii. We consider the scalar manifold of the supergravity
theory of the form $SO(8,8)/SO(8)\times SO(8)$ and find a number of
stable non-supersymmetric $AdS_3$ critical points for some values of
$\alpha$. These correspond to non-supersymmetric IR fixed points of
the UV $N=(4,4)$ SCFT dual to the maximally supersymmetric critical
point. We study the associated RG flow solutions interpolating
between these fixed points and the UV $N=(4,4)$ SCFT. Possible
supersymmetric flows to non-conformal field theories are also
investigated. Additionally, a half-supersymmetric domain wall within
this gauged supergravity is obtained.}
\begin{document}
\section{Introduction}
AdS$_3$/CFT$_2$ correspondence is interesting in various aspects.
Unlike in higher dimensional cases, much more insight to the AdS/CFT
correspondence \cite{maldacena} is expected since both gravity and
field theory sides are well under control. It is also useful in the
study of black hole entropy, see for example \cite{krause lecture}
and \cite{Strominger_lecture}. Until now, various gravity
backgrounds implementing AdS$_3$/CFT$_2$ correspondence have been
proposed. Some of them are obtained from Kaluza-Klein dimensional
reductions of higher dimensional supergravities on spheres or other
internal manifolds. The other are constructed directly within the
three dimensional framework of Chern-Simons gauged supergravity,
but, in some cases particularly for compact and non-compact gauge
groups, higher dimensional origins are still mysterious.
\\
\indent One of the most interesting backgrounds for AdS$_3$/CFT$_2$
correspondence is string theory on $AdS_3\times S^3\times S^3\times
S^1$. The background is half-supersymmetric and dual to large
$N=(4,4)$ SCFT in two dimensions, see \cite{Ali_classification} for
a classification of $N=4$ SCFT in two dimensions. In string theory,
this arises as a near horizon limit of the double D1-D5 brane system
\cite{AdS3S3S3S1_1, AdS3S3S3S1_2, AdS3S3S3S1_3}. The Kaluza-Klein
spectrum for small $S^1$ radius has been computed in
\cite{deBoer_largeN4}. Apart from the non-propagating supergravity
multiplet in three dimensions, the spectrum contains massive
multiplets of various spins. The full symmetry of $AdS_3\times
S^3\times S^3$ is $D^1(2,1;\alpha)\times D^1(2,1;\alpha)$ whose
bosonic subgroup is $SO(2,2)\times SO(4)\times SO(4)$ corresponding
to the isometry of $AdS_3\times S^3\times S^3$, respectively.
Additionally, the holography of large $N=4$ SCFT has recently been
studied in the context of higher spin $AdS_3$ dual
\cite{Gopakumar_LargeN4_dual}.
\\
\indent Like in higher dimensions, it would be useful to have an
effective theory in three dimensions that describes the above
$S^3\times S^3$ dimensional reduction. The $AdS_3\times S^3\times
S^3$ background will become an $AdS_3$ vacuum preserving sixteen
supercharges and $SO(4)\times SO(4)$ gauge symmetry, which is the
isometry of $S^3\times S^3$. This can be achieved by a gauged
matter-coupled supergravity in three dimensions \cite{nicolai1, N8,
dewit}. The gauge group should contain the $SO(4)\times SO(4)$
factor. The natural construction should be the $N=8$ gauged
supergravity since the number of supersymmetry is exactly the same
as that of the $AdS_3\times S^3\times S^3$ background. A theory
describing supergravity coupled to massive spin-$\frac{1}{2}$
multiplets has been studied in \cite{bs} in which some critical
points and a holographic RG flow have been discussed. The resulting
theory is in the form of $N=8$ gauged supergravity with compact
$SO(4)\times SO(4)$ gauge group and $SO(8,n)/SO(8)\times SO(n)$
scalar manifold.
\\
\indent When coupled to massive spin-1 multiplets, the theory needs
to accompany for massive vector fields. For a theory coupled to two
spin-1 multiplets, the corresponding gauge group is a non-semisimple
group $(SO(4)\times SO(4))\ltimes \mathbf{T}^{12}$. It has been
argued that the effective theory is the $N=8$ gauged supergravity
with $SO(8,8)/SO(8)\times SO(8)$ scalar manifold
\cite{Hohm_henning}. The gauging is a straightforward extension of
the $SO(4)\ltimes \mathbf{T}^6$ gauging of \cite{henning_N8AdS3_S3}
in which the effective theory of six-dimensional supergravity
reduced on $AdS_3\times S^3$ has been given. Some supersymmetric
vacua of the $(SO(4)\times SO(4))\ltimes \mathbf{T}^{12}$ gauged
theory have already been identified in \cite{gkn}. All of these
vacua are related to the maximally supersymmetric vacuum by marginal
deformations. The theory with only the $SO(4)\times SO(4)$
semisimple part of the gauge group being gauged has been study in
\cite{HS2}, and the solution corresponding to a marginal deformation
from $N=(4,4)$ to $N=(3,3)$ SCFT, describing a D5-brane
reconnection, has been explicitly given.
\\
\indent In this paper, we will reexamine the full $(SO(4)\times
SO(4))\ltimes \mathbf{T}^{12}$ gauging and look for other
deformations apart from the marginal ones. This could be relevant
for AdS$_3$/CFT$_2$ correspondence and black hole physics. The
holographic study of the conformal symmetry $D^1(2,1;\alpha)$ is not
only useful in the context of AdS$_3$/CFT$_2$ correspondence but
also in AdS$_2$/CFT$_1$ correspondence. This is because the symmetry
$D^1(2,1;\alpha)$ also arises in superconformal quantum mechanics
\cite{SCQ_1, SCQ_2, SCQ_3}. The isometry of $AdS_2$ is $SO(2,1)$
which is a subgroup of the $AdS_3$ isometry $SO(2,2)\sim
SO(2,1)\times SO(2,1)$. Accordingly, the superconformal symmetry in
one dimension contains only a single $D^1(2,1;\alpha)$. The
holographic study of AdS$_2$/CFT$_1$ correspondence directly from
two dimensional gauged supergravity has not been performed
extensively. This is in part due to the lack of gauged
supergravities in two dimensions. Until now, only the maximal gauged
supergravity and its truncation have appeared \cite{Henning_2D,
Henning_RotatingD0}. Since $AdS_2$ can be obtained by dimensional
reduction of $AdS_3$ on $S^1$ via a very-near-horizon limit
\cite{AdS2fromAdS3_1, AdS2fromAdS3_2}, the results obtained here
might be useful in the study of deformations in $D^1(2,1;\alpha)$
superconformal mechanics.
\\
\indent The paper is organized as follow. In section
\ref{N8_theory}, we will give a brief review of $N=8$, $(SO(4)\times
SO(4))\ltimes \mathbf{T}^{12}$ gauged supergravity along with some
relations to the $N=(4,4)$ SCFT. Section \ref{critical_point} deals
with a description of new critical points found in this work. In
section \ref{flow}, we will give some RG flow solutions describing
deformations of the $N=(4,4)$ SCFT to other SCFTs in the IR. The
possible supersymmetric flows to non-conformal theories are also
explored in this section. We end the paper by giving some
conclusions and discussions in section \ref{conclusion}. The
appendices summarize necessary ingredients needed in the
construction of $N=8$ theory and relevant formulae including the
explicit form of some scalar potentials.
\section{$N=8$, $(SO(4)\times SO(4))\ltimes \mathbf{T}^{12}$ gauged supergravity in three dimensions }\label{N8_theory}
We now review the construction of $N=8$ gauged supergravity with
$(SO(4)\times SO(4))\ltimes \mathbf{T}^{12}$ gauge group. The theory
has partially been studied before in \cite{gkn}. We will explore the
scalar potential of this theory in more details. Rather than follow
the parametrization of $SO(8,8)/SO(8)\times SO(8)$ coset manifold as
in \cite{gkn}, we will use the parametrization similar to that of
\cite{DW3D}. In this parametrization, it is more convenient to
determine the residual gauge symmetry while the parametrization used
in \cite{gkn} gives a simple action of the translation generators
$\mathbf{T}^{12}$ on scalar fields.
\\
\indent It has been argued in \cite{Hohm_henning} that this theory
is an effective theory of ten dimensional supergravity on
$AdS_3\times S^3\times S^3\times S^1$, or nine dimensional
supergravity on $AdS_3\times S^3\times S^3$ for small $S^1$ radius,
and describes the coupling of two massive spin-1 multiplets,
containing twelve vectors, to the non-propagating supergravity
multiplet of the reduction. All together, the resulting theory is
$N=8$ gauged supergravity with the scalar manifold
$SO(8,8)/SO(8)\times SO(8)$ and $(SO(4)\times SO(4))\ltimes
\mathbf{T}^{12}$ gauge group.
\\
\indent The whole construction is similar to that given in
\cite{gkn} and \cite{DW3D}. We will work in the $SO(8)$ R-symmetry
covariant formulation of \cite{dewit} with some relevant formulae
and details explicitly given in appendix \ref{detail}. We first
introduce the basis for a $GL(16,\mathbb{R})$ matrices
\begin{equation}
(e_{mn})_{pq}=\delta_{mp}\delta_{nq},\qquad m,n,p,q=1,\ldots,16\, .
\end{equation}
The compact generators of $SO(8,8)$ are then given by
\begin{eqnarray}
SO(8)^{(1)}&:&\qquad J_1^{IJ}=e_{JI}-e_{IJ},\qquad I,J =1,\ldots,
8,\nonumber \\
SO(8)^{(2)}&:&\qquad J_2^{rs}=e_{s+8,r+8}-e_{r+8,s+8} ,\qquad r,s
=1,\ldots, 8\, .
\end{eqnarray}
The non-compact generators corresponding to 64 scalars are
identified as
\begin{equation}
Y^{Kr}=e_{K,r+8}+e_{r+8,K},\qquad K,r =1,\ldots , 8\, .
\end{equation}
\indent In the formulation of \cite{dewit}, scalars transform as a
spinor under $SO(8)_R$ R-symmetry. It can be easily seen from the
above equation that $Y^{Kr}$ transform as a vector under $SO(8)_R$
identified with $SO(8)^{(1)}$ with generators $J^{IJ}_1$. We define
the following $SO(8)_R$ generators in a spinor representation by
\begin{equation}
T^{IJ}=\left(
         \begin{array}{cc}
           \Gamma^{IJ} & 0 \\
           0 & 0 \\
         \end{array}
       \right)
\end{equation}
constructed from the $8\times 8$ $SO(8)$ gamma matrices $\Gamma^I$.
We have defined
\begin{equation}
\Gamma^{IJ}=-\frac{1}{4}\left(\Gamma^I(\Gamma^J)^T-\Gamma^J(\Gamma^I)^T\right)
\end{equation}
with the $8\times 8$ gamma matrices $\Gamma^I$ are given in appendix
\ref{detail}.
\\
\indent The gauge group $(SO(4)\times SO(4))\ltimes \mathbf{T}^{12}$
is embedded in $SO(8,8)$ as follow. We first form a diagonal
subgroup of $SO(8)\times SO(8)$ with generators
\begin{equation}
SO(8)_{\textrm{diag}}:\qquad J^{AB}=J_1^{AB}+J^{AB}_2,\qquad
A,B=1,\ldots, 8\, .
\end{equation}
The $SO(4)\times SO(4)$ part is generated by
\begin{eqnarray}
SO(4)^+:& &\qquad j_1^{ab}=J^{ab}, \nonumber \\
SO(4)^-:& &\qquad
j_2^{\hat{a}\hat{b}}=J^{\hat{a}+4,\hat{b}+4},\qquad
a,b,\hat{a},\hat{b} =1,\ldots, 4\, .
\end{eqnarray}
The ``hat'' indices refer to $SO(4)^-$. We now construct the
translational generators $\mathbf{T}^{28}$ as in \cite{DW3D}
\begin{equation}
t^{AB}=J_1^{AB}-J_2^{AB}+Y^{BA}-Y^{AB}
\end{equation}
and identify $\mathbf{T}^{12}\sim\mathbf{T}^6\times \mathbf{T}^6$
generators as
\begin{eqnarray}
t_1^{ab}=t^{ab},\qquad
t_2^{\hat{a}\hat{b}}=t^{\hat{a}+4,\hat{b}+4},\qquad
a,b,\hat{a},\hat{b} =1,\ldots, 4\, .
\end{eqnarray}
\indent The gauge group is embedded in $SO(8,8)$ with a specific
form of the embedding tensor. As shown in \cite{csym}, there is no
coupling among the $SO(4)^{\pm}$. The gauging is very similar to the
$SO(4)\ltimes \mathbf{T}^6$ gauged supergravity constructed in
\cite{henning_N8AdS3_S3} with two factors of $SO(4)\ltimes
\mathbf{T}^6$. The embedding tensor is simply given by two copies of
that given in \cite{henning_N8AdS3_S3}. We end up with two
independent coupling constants
\begin{equation}
\Theta = g_1\Theta_1+g_2\Theta_2\, .
\end{equation}
where $\Theta_{1,2}$ describe the embedding of each $SO(4)\ltimes
\mathbf{T}^6$ factor of the full gauge group. We should note that
supersymmetry allows for four independent couplings namely between
the moment maps $g'_1(\mc{V}(j_1^{ab}),\mc{V}(t_1^{ab}))$,
$g'_2(\mc{V}(t_1^{ab}),\mc{V}(t_1^{ab}))$,
$g'_3(\mc{V}(j_2^{ab}),\mc{V}(t_2^{ab}))$ and
$g'_4(\mc{V}(t_2^{ab}),\mc{V}(t_2^{ab}))$ in the T-tensor, see
\cite{henning_N8AdS3_S3} and \cite{gkn}. We have used a shorthand
notation for $\mc{V}^{\mc{M}}_{\phantom{as}\mc{A}}$. However, the
requirement that the theory admits a maximally supersymmetric vacuum
at the origin of the scalar manifold imposes two conditions on the
original four couplings. In more detail, the two conditions require
$g'_2=-g'_1$ and $g'_4=-g'_3$. After rename the relevant couplings,
we end up with the embedding tensor
\begin{equation}
\Theta_{abcd} =
g_1\epsilon_{abcd}^++g_2\epsilon_{\hat{a}\hat{b}\hat{c}\hat{d}}^-\,
.
\end{equation}
\indent This embedding tensor together with the formulae in appendix
\ref{detail} and an explicit parametrization of the coset
representative of $SO(8,8)/SO(8)\times SO(8)$ can be used to compute
the scalar potential. We will analyze the resulting potential on
submanifolds of $SO(8,8)/SO(8)\times SO(8)$ invariant under some
subgroups of $SO(4)\times SO(4)$ in the next section.
\\
\indent Before looking at the critical points, we give a review of
the relation between $(SO(4)\times SO(4))\ltimes \mathbf{T}^{12}$,
$N=8$ gauged supergravity and $N=(4,4)$ SCFT. The semisimple part of
the gauge group $SO(4)^+\times SO(4)^-$ corresponds to the isometry
of $S^3\times S^3$. Together with the usual $SO(2,2)$ isometry of
$AdS_3$, they constitute the bosonic subgroup $SO(2,1)_L\times
SU(2)^+_L\times SU(2)^-_L\times SO(2,1)_R\times SU(2)^+_R\times
SU(2)^-_R$ of the superconformal group $D^1(2,1;\alpha)\times
D^1(2,1;\alpha)$ via the isomorphisms $SO(2,2)\sim SO(2,1)_L\times
SO(2,1)_R$ and $SO(4)^\pm \sim SU(2)^\pm_L\times SU(2)^\pm_R$. The
$\alpha$ parameter is identified with the ratio of the coupling
constant $g_2=\alpha g_1$. For positive $\alpha$, the theory
describes the dimensional reduction of nine dimensional supergravity
on $S^3\times S^3$. For negative $\alpha$, it may possibly describe
the reduction on $S^3\times H^3$ where $H^3$ is a hyperbolic space
in three dimensions.
\\
\indent The translational part $\mathbf{T}^{12}$ of the gauge group
describes twelve massive vector fields \cite{csym}. The massive
vector fields will show up in the vacuum of the theory via twelve
massless scalars in the adjoint representation of $SO(4)\times
SO(4)$. These are Goldstone bosons for the $\mathbf{T}^{12}$
symmetry since the vacuum is invariant only under $SO(4)^+\times
SO(4)^-$ not the full gauge group. We will see this when we compute
the mass spectrum of scalar fields.
\section{Some critical points of $N=8$, $(SO(4)\ltimes SO(4))\ltimes \mathbf{T}^{12}$ gauged supergravity}\label{critical_point}
We now look for critical points of the $N=8$ gauged supergravity
constructed in the previous section. Analyzing the scalar potential
on the full 64-dimensional scalar manifold $SO(8,8)/SO(8)\times
SO(8)$ is beyond our reach with the present-time computer. We then
employ an effective method given in \cite{warner} to find some
interesting critical points on a submanifold invariant under some
subgroup of the gauge group. A group theoretical argument guarantees
that the corresponding critical points are critical points of the
scalar potential on the full scalar manifold. Even on these
truncated manifolds, the explicit form of the potential is still
very complicated. Therefore, in most cases, we refrain from giving
the full expression for the potential.
\\
\indent At the trivial critical point with all scalars vanishing,
the full gauge group $(SO(4)\times SO(4))\ltimes \mathbf{T}^{12}$ is
broken down to its maximal compact subgroup $SO(4)\times SO(4)$
corresponding to the isometry of $S^3\times S^3$. The 64 scalars
transform under $SO(8)\times SO(8)\subset SO(8,8)$ as
$(\mathbf{8},\mathbf{8})$. Then, under the $SO(4)^+\times
SO(4)^-\subset SO(8)_{\textrm{diag}}$, they transform as
\begin{eqnarray}
\mathbf{8}\times
\mathbf{8}&=&\left[(\mathbf{4}^+,\mathbf{1}^+)+(\mathbf{1}^-,\mathbf{4}^-)\right]\times
\left[(\mathbf{4}^+,\mathbf{1}^+)+(\mathbf{1}^-,\mathbf{4}^-)\right]\nonumber
\\
&=&(\mathbf{1}^++\mathbf{6}^++\mathbf{9}^+,\mathbf{1}^+)+(\mathbf{1}^-,\mathbf{1}^-+\mathbf{6}^-+\mathbf{9}^-)+(\mathbf{4}^+,\mathbf{4}^-)
+(\mathbf{4}^-,\mathbf{4}^+).\quad \, \, \, \label{scalar_decom1}
\end{eqnarray}
We can further decompose the above representations into
$SU(2)^+_L\times SU(2)^+_R\times SU(2)^-_L\times SU(2)^-_R$
representations labeled by $(\ell^+_L,\ell^+_R;\ell^-_L,\ell^-_R)$
as follow:
\begin{eqnarray}
\mathbf{8}\times
\mathbf{8}&=&(\mathbf{1},\mathbf{1};\mathbf{1},\mathbf{1})+(\mathbf{1},\mathbf{3};\mathbf{1},\mathbf{1})
+(\mathbf{3},\mathbf{1};\mathbf{1},\mathbf{1})+(\mathbf{3},\mathbf{3};\mathbf{1},\mathbf{1})\nonumber
\\
& &
+(\mathbf{1},\mathbf{1};\mathbf{1},\mathbf{1})+(\mathbf{1},\mathbf{1};\mathbf{1},\mathbf{3})
+(\mathbf{1},\mathbf{1};\mathbf{3},\mathbf{1})+(\mathbf{1},\mathbf{1};\mathbf{3},\mathbf{3})\nonumber
\\
& &
+(\mathbf{2},\mathbf{2};\mathbf{2},\mathbf{2})+(\mathbf{2},\mathbf{2};\mathbf{2},\mathbf{2}).\label{scalar_decom2}
\end{eqnarray}
The result precisely agrees with the representation content obtained
from the $AdS_3\times S^3\times S^3$ reduction
\cite{deBoer_largeN4}. For conveniences, we also repeat the massive
spin-1 supermultiplets $(0,1;0,1)_{\rm S}$ and $(1,0;1,0)_{\rm S}$
of the $AdS_3\times S^3\times S^3$ reduction in Table \ref{spin11}
and \ref{spin12}.
\begin{table}[bt]
\centering
  \begin{tabular}{|c|c|c|c|}
  \hline
   \raisebox{-1.25ex}{$h_L$} \raisebox{1.25ex}{$h_R$} &
   $\frac{\alpha}{1+\alpha}$ & $\frac{3\alpha+1}{2(1+\alpha)}$
   & $\frac{2\alpha+1}{1+\alpha}$
   \rule[-2ex]{0pt}{5.5ex}\\
    \hline
   $\frac{\alpha}{1+\alpha}$ & $(0,1;0,1)$  & $(0,1;\frac{1}{2},\frac{1}{2})$ & $(0,1;0,0)$
   \rule[-1.5ex]{0pt}{4ex}\\
    \hline
   $\frac{3\alpha+1}{2(1+\alpha)}$ &  $(\frac{1}{2},\frac{1}{2};0,1)$
   & $(\frac{1}{2},\frac{1}{2};\frac{1}{2},\frac{1}{2})$ &
   $(\frac{1}{2},\frac{1}{2};0,0)$  \rule[-1.5ex]{0pt}{4ex} \\
     \hline
   $\frac{2\alpha+1}{1+\alpha}$ &  $(0,0;0,1)$ & $(0,0;\frac{1}{2},\frac{1}{2})$
   & $(0,0,0,0)$ \rule[-1.5ex]{0pt}{4ex} \\
  \hline
  \end{tabular}
      \caption{The massive spin-1 multiplet $(0,1;0,1)_{\rm S}$.}
\label{spin11}
\end{table}
\begin{table}[bt]
\centering
  \begin{tabular}{|c|c|c|c|}
  \hline
   \raisebox{-1.25ex}{$h_L$} \raisebox{1.25ex}{$h_R$} &
   $\frac{1}{1+\alpha}$ & $\frac{3+\alpha}{2(1+\alpha)}$
   & $\frac{2+\alpha}{1+\alpha}$
   \rule[-2ex]{0pt}{5.5ex}\\
    \hline
   $\frac{1}{1+\alpha}$ & $(1,0;1,0)$  & $(1,0;\frac{1}{2},\frac{1}{2})$ & $(1,0;0,0)$
   \rule[-1.5ex]{0pt}{4ex}\\
    \hline
   $\frac{3+\alpha}{2(1+\alpha)}$ &  $(\frac{1}{2},\frac{1}{2};1,0)$
   & $(\frac{1}{2},\frac{1}{2};\frac{1}{2},\frac{1}{2})$ &
   $(\frac{1}{2},\frac{1}{2};0,0)$  \rule[-1.5ex]{0pt}{4ex} \\
     \hline
   $\frac{2+\alpha}{1+\alpha}$ &  $(0,0;1,0)$ & $(0,0;\frac{1}{2},\frac{1}{2})$
   & $(0,0,0,0)$ \rule[-1.5ex]{0pt}{4ex} \\
  \hline
  \end{tabular}
      \caption{The massive spin-1 multiplet $(1,0;1,0)_{\rm S}$.}
\label{spin12}
\end{table}
\\
\indent We can now compute the scalar potential by using the
formulae in appendix \ref{detail}. After expanding the potential
around $L=\mathbf{I}$, we find the scalar mass spectrum at the
maximally supersymmetric vacuum as shown in Table \ref{spectrum1}.
\begin{table}[bt]
\centering
\begin{tabular}{|c|c|}
  \hline
  $SO(4)^+\times SO(4)^-$ & $m^2L^2$ \\ \hline
  $(\mathbf{1},\mathbf{1})$ & $\frac{4g_1(2g_1+g_2)}{(g_1+g_2)^2}$ \\
  $(\mathbf{6},\mathbf{1})$ & $0$ \\
  $(\mathbf{9},\mathbf{1})$ & $-\frac{4g_1g_2}{(g_1+g_2)^2}$ \\
  $(\mathbf{1},\mathbf{1})$ & $\frac{4g_2(2g_2+g_1)}{(g_1+g_2)^2}$  \\
  $(\mathbf{1},\mathbf{6})$ & $0$ \\
  $(\mathbf{1},\mathbf{9})$ & $-\frac{4g_1g_2}{(g_1+g_2)^2}$ \\
  $(\mathbf{4},\mathbf{4})$ & $\frac{3g_2^2-2g_1g_2-g_1^2}{(g_1+g_2)^2}$ \\
  $(\mathbf{4},\mathbf{4})$ & $\frac{3g_1^2-2g_1g_2-g_2^2}{(g_1+g_2)^2}$ \\
  \hline
\end{tabular}
      \caption{The mass spectrum of the trivial critical point.}
\label{spectrum1}
\end{table}
The $AdS_3$ radius is given by $L=\frac{1}{\sqrt{-V_0}}$, and the
value of the potential at this point is $V_0=-64(g_1+g_2)^2$. Using
the relation $m^2L^2=\Delta (\Delta-2)$ and $\Delta=h_L+h_R$, we can
verify that the mass spectrum agrees with the values of $h_R$ and
$h_L$ in Table \ref{spin11} and \ref{spin12}. As mentioned before,
there are twelve massless Goldstone bosons transforming in the
adjoint representation
$(\mathbf{1},\mathbf{6})+(\mathbf{6},\mathbf{1})$ of $SO(4)\times
SO(4)$. Note also that there is a Minkowski vacuum at $g_2=-g_1$ or
$\alpha=-1$.
\subsection{Critical points on the $SO(4)_{\textrm{diag}}$ invariant manifold}
We first consider scalars which are singlets under the diagonal
subgroup $SO(4)_{\textrm{diag}}\subset SO(4)\times SO(4)$. To obtain
representations of the scalars under this subgroup, we take a tensor
product in the last line of \eqref{scalar_decom1}. We find that
there are four singlets, two from the obvious ones
$(\mathbf{1}^+\times \mathbf{1}^+,\mathbf{1}^-\times \mathbf{1}^-)$
and the other two from the product $(\mathbf{4}^+\times
\mathbf{4}^-,\mathbf{4}^-\times \mathbf{4}^+)$. They correspond to
the following non-compact generators
\begin{eqnarray}
\tilde{Y}_1&=&Y^{11}+Y^{22}+Y^{33}+Y^{44},\qquad
\tilde{Y}_2=Y^{55}+Y^{66}+Y^{77}+Y^{88},\nonumber \\
\tilde{Y}_3&=&Y^{51}+Y^{62}+Y^{73}+Y^{84},\qquad
\tilde{Y}_4=Y^{15}+Y^{26}+Y^{37}+Y^{48}\, .
\end{eqnarray}
The coset representative is accordingly parametrized by
\begin{equation}
L=e^{a_1\tilde{Y}_1}e^{a_2\tilde{Y}_2}e^{a_3\tilde{Y}_3}e^{a_4\tilde{Y}_4}\,
.\label{SO4d_coset}
\end{equation}
Apart from the trivial critical point at $a_1=a_2=a_3=a_4=0$, we
find the following critical points.
\begin{itemize}
  \item A $dS_3$ critical point is at $a_1=\frac{1}{2}\ln
  \frac{g_2}{g_2-g_1}$, $a_2=\frac{1}{2}\ln
  \frac{g_1}{g_1-g_2}$ and $a_3=a_4=0$. The cosmological constant is
  $V_0=\frac{64g_1^2g_2^2}{(g_1-g_2)^2}$. This critical point has $SO(4)\times
  SO(4)$ symmetry since $a_3=a_4=0$. Non-zero $a_3$ or $a_4$ would break $SO(4)\times
  SO(4)$ to its diagonal subgroup.
  \item A non-supersymmetric $AdS_3$ is given by $a_1=\frac{1}{2}\ln
  \frac{\sqrt{g_1-4g_2}-\sqrt{g_1}}{2\sqrt{g_1}}$ and $a_2=a_3=a_4=0$. The cosmological
  constant is
  \begin{equation}
V_0=-32\left[g_1^2+4g_2^2-6g_1g_2+(4g_2-g_1)\sqrt{g_1(g_1-4g_2)}\right].
  \end{equation}
  $a_1$ is real for $g_1>0$ and $g_2<0$, and the critical point is
  $AdS_3$, $V_0<0$, for $g_1>0$ and $g_2<-\frac{\sqrt{2}+1}{2}g_1$.
  An equivalent critical point is given by $a_2\neq 0$ and
  $a_1=a_3=a_4=0$ but with $g_1\leftrightarrow g_2$. For later
  reference, we will call this critical point $P_1$.
\item Another non-supersymmetric critical point is at $a_4=\ln
  \frac{\sqrt{g_1}+\sqrt{3g_2}}{\sqrt{g_1}-\sqrt{3g_2}}$ with
  $g_2=\frac{1}{9}\left(\sqrt{13}-2\right)g_1$ and
  $V_0=-\frac{8}{3}\left(43+13\sqrt{13}\right)g_1^2$. In this case,
  only a specific value of $\alpha$ gives a critical point. The
  residual gauge symmetry in this case is $SO(4)_{\textrm{diag}}$. We
  will label this critical point as $P_2$.
  \item There is another $dS_3$ critical point which is invariant
under $SO(4)\times SO(4)$ and characterized by
\begin{eqnarray}
a_1&=&\frac{1}{2}\ln \frac{g_2}{g_2-g_1},\qquad a_2=\frac{1}{2}\ln
\frac{g_1-2g_2}{g_1-g_2},\nonumber\\
V_0&=&\frac{64g_2^2\left(g_1^3-g_1^2g_2-8g_2^3\right)}{(g_1-g_2)^3}\,
.
\end{eqnarray}
$a_1$ and $a_2$ are real for $g_1<0$ and $g_2<g_1$. In this range,
we find $V_0>0$, so this critical point is $dS_3$.
\end{itemize}
The full scalar potential for the four scalars is given in appendix
\ref{potential_form}.
\\
\indent At this stage, we should note an interesting result
discovered in \cite{HS2} but with a compact gauge group $SO(4)\times
SO(4)$. This solution describes a marginal deformation of $N=(4,4)$
SCFT to $N=(3,3)$ SCFT and has an interpretation in term of a
reconnection of D5-branes in the double D1-D5 system. The solution
is also encoded in our present framework. In this case, we must set
$g_2=g_1$, or equivalently setting $\alpha=1$ in order to get
massless (marginal) scalars preserving the $SO(4)$ diagonal subgroup
of $SO(4)\times SO(4)$.
\\
\indent Follow \cite{HS2}, we further truncate the four scalars to
two via
\begin{equation}
a_2=a_1,\qquad a_4=-a_3\, .
\end{equation}
This is a consistent truncation for $g_2=g_1$ since it corresponds
to a fixed point of an inner automorphism that leaves the embedding
tensor invariant \cite{HS2}. We find a critical point at
\begin{equation}
e^{a_1+a_3}=1+\sqrt{1-e^{2a_1}},\qquad V_0=-256g_1^2
\end{equation}
with the corresponding $A_1$ tensor given by
\begin{equation}
A_1^{IJ}=\textrm{diag}\left(-8g_1,-8g_1,-8g_1,8g_1,8g_1,8g_1,-8g_1\sqrt{4e^{-2a_1}-3},8g_1\sqrt{4e^{-2a_1}-3}\right).
\end{equation}
We can see that as long as $a_1\neq 0$, the $N=(4,4)$ supersymmetry
is broken to $N=(3,3)$. We refer the reader to \cite{HS2} for the
full discussion of this vacuum.
\\
\indent We now analyze the scalar masses at critical points $P_1$ and $P_2$ to check their stability. For critical point $P_1$,
it is useful to classify
the 64 scalars according to their representations under the residual
symmetry $SO(4)\times SO(4)$. The result is shown in Table
\ref{spectrum2}.
\begin{table}[bt]
\centering
\begin{tabular}{|c|c|}
  \hline
  $SO(4)^+\times SO(4)^-$ & $m^2L^2$ \\ \hline
  $(\mathbf{1},\mathbf{1})$ & $\frac{12g_2}{g_2+\sqrt{g_1(g_1-4g_2)}}$ \\
                                          & $-\frac{16g_2^2+20g_1g_2-6g_1^2+2(g_1+2g_2)\sqrt{g_1(g_1-4g_2)}}
                                          {g_1^2-4g_1g_2-4g_2^2}$ \\
                                          &  $\frac{4g_2^2+14g_1g_2-3g_1^2+(4g_2-g_1)\sqrt{g_1(g_1-4g_2)}}{2(g_1^2-4g_1g_2-4g_2^2)}$   \\
                                          &   $-\frac{3g_1^2-30g_1g_2+12g_2^2+3(3g_1-4g_2)\sqrt{g_1(g_1-4g_2)}}{2(g_1^2-4g_1g_2-4g_2^2)}$  \\
  $(\mathbf{6},\mathbf{1})$ & $0$ \\
  $(\mathbf{9},\mathbf{1})$ & $\frac{8g_1g_2}{g_1^2-6g_1g_2+(2g_2-g_1)\sqrt{g_1(g_1-4g_2)}}$ \\
  $(\mathbf{1},\mathbf{1})$ & $\frac{4g_2(2g_2+g_1)}{(g_1+g_2)^2}$  \\
  $(\mathbf{1},\mathbf{6})$ & $0$ \\
  $(\mathbf{1},\mathbf{9})$ & $-\frac{4g_1^2-24g_1g_2-8g_2^2+4(g_1-g_2)\sqrt{g_1(g_1-4g_2)}}{g_1^2-4g_1g_2-4g_2^2}$ \\
  $(\mathbf{4},\mathbf{4})$ & $\frac{4g_2^2+14g_1g_2-3g_1^2+(4g_2-g_1)\sqrt{g_1(g_1-4g_2)}}{2(g_1^2-4g_1g_2-g_2^2)}$ \\
  $(\mathbf{4},\mathbf{4})$ & $-\frac{12g_2^2-30g_1g_2+3g_1^2+(9g_1-12g_2)\sqrt{g_1(g_1-4g_2)}}{2(g_1^2-4g_1g_2-g_2^2)}$ \\
  \hline
\end{tabular}
      \caption{The scalar mass spectrum of the $SO(4)\times SO(4)$ critical point $P_1$.}
\label{spectrum2}
\end{table}
Similar to the trivial critical point, there are 12 massless scalars
corresponding to the broken $\mathbf{T}^{12}$ symmetry. The
stability bound, or BF bound $m^2L^2\geq -1$, is satisfied by
$-\frac{13+9\sqrt{2}}{2}g_1<g_2<-\frac{1+\sqrt{1+\sqrt{2}}}{2}g_1$.
\\
\indent For critical point $P_2$, we can compute all scalar masses as shown in Table
\ref{spectrum3}. It is easily seen that all masses satisfy the BF
bound. There are 18 massless Goldstone bosons corresponding to the
symmetry breaking $(SO(4)\times SO(4))\ltimes
\mathbf{T}^{12}\rightarrow SO(4)$.
\begin{table}[bt]
\centering
\begin{tabular}{|c|c|}
  \hline
  $SO(4)$ & $m^2L^2$ \\ \hline
  $\mathbf{1}$ &$13.6358$, $6.0931$, $3.3703$, $3.1180$\\
  $\mathbf{6}$ & $0_{(\times 18)}$ \\
  $\mathbf{9}$ & $\frac{8}{29}(7\sqrt{13}-12)_{(\times 9)}$, $\frac{4}{29}(5\sqrt{13}-21)_{(\times 9)}$,\\
                       & $\frac{4}{29}(8+5\sqrt{13})_{(\times 9)}$, $\frac{4}{87}(19\sqrt{13}-74)_{(\times 9)}$ \\
  \hline
\end{tabular}
      \caption{The scalar mass spectrum of the $SO(4)$ critical point $P_2$ for $g_2=\frac{\sqrt{13}-2}{9}g_1$.}
\label{spectrum3}
\end{table}

\subsection{Critical points on the $SO(2)_{\textrm{diag}}\times SO(2)_{\textrm{diag}}$ invariant manifold}
We now proceed to consider a smaller residual symmetry
$SO(2)_{\textrm{diag}}\times SO(2)_{\textrm{diag}}\subset
SO(4)_{\textrm{diag}}$. Under $SO(2)\times SO(2)$, the $SO(4)$
fundamental representation $\mathbf{4}$ decomposes according to
$\mathbf{4}\rightarrow
(\mathbf{2},\mathbf{1})+(\mathbf{1},\mathbf{2})$. Substituting this
decomposition for $\mathbf{4}^+$ and $\mathbf{4}^-$ in
\eqref{scalar_decom1} and taking the product to form a diagonal
subgroup, we find that there are sixteen singlets given by the
non-compact generators
\begin{eqnarray}
\bar{Y}_1&=&Y^{11}+Y^{22},\qquad \bar{Y}_2=Y^{33}+Y^{44},\qquad
\bar{Y}_3=Y^{55}+Y^{66},\qquad \bar{Y}_4=Y^{77}+Y^{88},\nonumber \\
\bar{Y}_5&=&Y^{15}+Y^{26},\qquad \bar{Y}_6=Y^{37}+Y^{48},\qquad
\bar{Y}_7=Y^{51}+Y^{62},\qquad \bar{Y}_8=Y^{73}+Y^{84},\nonumber \\
\bar{Y}_9&=&Y^{12}-Y^{21},\qquad \bar{Y}_{10}=Y^{34}-Y^{43},\qquad
\bar{Y}_{11}=Y^{56}-Y^{65},\qquad \bar{Y}_{12}=Y^{78}-Y^{87},\nonumber \\
\bar{Y}_{13}&=&Y^{16}-Y^{25},\qquad
\bar{Y}_{14}=Y^{38}-Y^{47},\qquad \bar{Y}_{15}=Y^{52}-Y^{61},\qquad
\bar{Y}_{16}=Y^{74}-Y^{83}\, .\nonumber \\
& &
\end{eqnarray}
The coset representative can be parametrized by
\begin{equation}
L=\prod_{i=1}^{16} e^{a_i\bar{Y}_i}\, .\label{SO2_SO2_coset}
\end{equation}
\indent Unlike the previous case, the scalar potential is so
complicated that it is not possible to make the full analysis.
However, with some ansatz, we find one non-trivial critical point at
\begin{eqnarray}
a_1&=&a_2=\frac{1}{2}\ln 2,\qquad a_3=-a_4=\frac{1}{2}\ln
\frac{g_2-6g_1+\sqrt{36g_1^2-12g_1g_2-3g_2^2}}{2g_2},\nonumber
\\
V_0&=&64(8g_1^2-g_2^2).
\end{eqnarray}
$a_3$ and $a_4$ are real for $g_1>0$ and $g_2\geq -6g_1$. In this
range, we find $V_0<0$ if $g_2<-2\sqrt{2}g_1$. Therefore, it is
possible to have an $AdS_3$ critical point. The residual symmetry is
$SO(4)\times SO(2)\times SO(2)$. We will denote this critical point
by $P_3$ for later reference.
\\
\indent
The stability of this critical point
can be verified from the scalar mass spectrum given in Table \ref{spectrum4}
\begin{table}[bt]
\centering
\begin{tabular}{|c|c|}
  \hline
  $SO(4)\times SO(2)\times SO(2)$ & $m^2L^2$ \\ \hline
  $(\mathbf{4},\mathbf{2},\mathbf{1})$ & $-\frac{60g_1^2-14g_1g_2+g_2^2+(6g_1-3g_2)\sqrt{36g_1^2-12g_1g_2-3g_2^2}}{16g_1^2-2g_2^2}$ \\
  $(\mathbf{4},\mathbf{1},\mathbf{2})$ & $-\frac{60g_1^2-24g_1g_2+g_2^2+(3g_2-6g_1)\sqrt{36g_1^2-12g_1g_2-3g_2^2}}{16g_1^2-2g_2^2}$ \\
  $(\mathbf{4},\mathbf{2},\mathbf{1})$ & $-\frac{124g_1^2-3g_2^2+(g_2+6g_1)\sqrt{36g_1^2-12g_1g_2-3g_2^2}}{16g_1^2-2g_2^2}$ \\
  $(\mathbf{4},\mathbf{1},\mathbf{2})$ & $-\frac{124g_1^2-3g_2^2-(g_2+6g_1)\sqrt{36g_1^2-12g_1g_2-3g_2^2}}{16g_1^2-2g_2^2}$  \\
  $(\mathbf{1},\mathbf{2},\mathbf{1})$ & $\frac{6g_2^2+24g_1g_2-72g_1^2+2(g_2-6g_1)\sqrt{36g_1^2-12g_1g_2-3g_2^2}}{8g_1^2-g_2^2}$ \\
  $(\mathbf{1},\mathbf{1},\mathbf{2})$ & $\frac{6g_2^2+24g_1g_2-72g_1^2-2(g_2-6g_1)\sqrt{36g_1^2-12g_1g_2-3g_2^2}}{8g_1^2-g_2^2}$ \\
  $(\mathbf{9},\mathbf{1},\mathbf{1})$ & $\frac{48g_1^2}{g_2^2-8g_1^2}$ \\
  $(\mathbf{6},\mathbf{1},\mathbf{1})$ & $0$ \\
  $2\times (\mathbf{1},\mathbf{2},\mathbf{2})$ & $0$ \\
  $2\times (\mathbf{1},\mathbf{1},\mathbf{1})$ & $0$ \\
  $(\mathbf{1},\mathbf{1},\mathbf{1})$ & $\alpha_1$, $\alpha_2$, $\alpha_3$ \\
  \hline
\end{tabular}
      \caption{The scalar mass spectrum of the $SO(4)\times SO(2)\times SO(2)$ critical point $P_3$.}
\label{spectrum4}
\end{table}
in which $\alpha_i$ are eigenvalues of the submatrix
\begin{equation}
\frac{1}{8g_1^2-g_2^2}\left(
  \begin{array}{ccc}
    -80g_1^2 & x_1 & x_2 \\
    x_1 & -\frac{g_2^2}{3} & -\frac{2g_2^2}{3} \\
    x_2 & -\frac{2g_2^2}{3} & -\frac{g_2^2}{3} \\
  \end{array}
\right)
\end{equation}
with the following elements
\begin{eqnarray}
x_1&=&2\sqrt{2}g_1\left(6g_1+g_2-\sqrt{36g_1^2-12g_1g_2-3g_2^2}\right)\nonumber \\
\textrm{and}\qquad
x_2&=&2\sqrt{2}g_1\left(6g_1+g_2+\sqrt{36g_1^2-12g_1g_2-3g_2^2}\right).
\end{eqnarray}
Their numerical values can be obtained upon specifying the values of
$g_1$ and $g_2$.
\\
\indent For all but $(\mathbf{1},\mathbf{1},\mathbf{1})$ and
$(\mathbf{1},\mathbf{1},\mathbf{2})$ scalars, the masses are above
the BF bound for $-6g_1<g_2<-2\sqrt{2}g_1$. The mass squares of
$(\mathbf{1},\mathbf{1},\mathbf{1})$ scalars are above the BF bound
for $-6g_1<g_2<-4.47g_1$. For $(\mathbf{1},\mathbf{1},\mathbf{2})$
scalars, the mass squares are above the BF bound for
$-6g_1<g_2<\mc{X}$ with $\mc{X}$ being the first root of
$p(\mc{X})=1088g_1^4-384g_1^3\mc{X}
+352g_1^2\mc{X}^2-144g_1\mc{X}^3-37\mc{X}^4=0$. This can be
translated to the value of $\alpha$ by setting $\mc{X}=\alpha g_1$.
The equation $p(\mc{X})=0$ gives the value of $\alpha=-5.93479$. The
stability is obtained in the range $-6g_1<g_2<-5.93479g_1$ which is
very narrow. Notice that for $g_2=-6g_1$, we find $a_3=a_4=0$, and
the symmetry is enhanced to $SO(4)\times SO(4)$. It can be checked
that this critical point indeed becomes critical point $P_1$ with
$g_2=-6g_1$.

\subsection{Critical points on the $SU(2)_L^+\times SU(2)_L^-$ invariant manifold}
One interesting deformation of $N=(4,4)$ SCFT is the chiral
supersymmetry breaking $(4,4)\rightarrow (4,0)$. The realization of
this breaking in the D1-D5 system has been studied in
\cite{Morales}. Another gravity dual of $N=(4,0)$ SCFT from string
theory has been studied in \cite{N04_1}, and the marginal
perturbation driving $N=(4,4)$ SCFT to the $N=(4,0)$ SCFT has been
identified in \cite{N04_2}. This supersymmetry breaking is not
possible in the compact $SO(4)\times SO(4)$ gauging of \cite{bs}
since there are no scalars which are singlets under a non-trivial
subgroup of $SO(4)\times SO(4)$ in order to become the R-symmetry of
$N=(4,0)$.
\\
\indent This is however possible in the present gauging. According
to \eqref{scalar_decom2}, we see that there are eight singlets under
$SU(2)_L^+\times SU(2)^-_L$ given by
\begin{equation}
(\mathbf{1},\mathbf{1};\mathbf{1},\mathbf{1})+(\mathbf{1},\mathbf{1};\mathbf{1},\mathbf{1})
+(\mathbf{1},\mathbf{3};\mathbf{1},\mathbf{1})+(\mathbf{1},\mathbf{1};\mathbf{1},\mathbf{3}).
\end{equation}
They correspond to the following non-compact generators
\begin{eqnarray}
\hat{Y}_1&=&Y^{11}+Y^{22}+Y^{33}+Y^{44},\qquad
\hat{Y}_2=Y^{12}-Y^{21}+Y^{34}-Y^{43},\nonumber \\
\hat{Y}_3&=&Y^{13}-Y^{31}-Y^{24}+Y^{42},\qquad
\hat{Y}_4=Y^{14}-Y^{41}+Y^{23}-Y^{32},\nonumber \\
\hat{Y}_5&=&Y^{55}+Y^{66}+Y^{77}+Y^{88},\qquad
\hat{Y}_6=Y^{56}-Y^{65}+Y^{78}-Y^{87},\nonumber \\
\hat{Y}_7&=&Y^{57}-Y^{75}-Y^{68}+Y^{86},\qquad
\hat{Y}_8=Y^{58}-Y^{85}+Y^{67}-Y^{76}\, .
\end{eqnarray}
We can parametrize the coset representative accordingly
\begin{equation}
L=e^{b_1\hat{Y}_1}e^{a_2\hat{Y}_2}e^{a_3\hat{Y}_3}e^{a_4\hat{Y}_4}e^{b_5\hat{Y}_5}
e^{a_6\hat{Y}_6}e^{a_7\hat{Y}_7}e^{a_8\hat{Y}_8}\label{L_SU2_SU2}
\end{equation}
in which $b_1$ and $b_5$ denote the $SO(4)\times SO(4)$ singlets.
Some critical points are given below.
\begin{itemize}
\item There is an $AdS_3$ critical point characterized by
\begin{eqnarray}
a_2&=&\cosh
^{-1}\sqrt{\frac{g_1+\sqrt{g_1(g_1-4g_2)}}{4g_1}},\nonumber \\
V_0&=&-32\left[g_1^2+4g_2^2-6g_1g_2+(4g_2-g_1)\sqrt{g_1(g_1-4g_2)}\right].
\end{eqnarray}
The cosmological constant is the same as $P_1$, but the residual
gauge symmetry is just $SO(4)^-\times SU(2)^+_L\times U(1)^+_R$ in
which $U(1)^+_R\subset SU(2)^+_R$.
\item There is a $dS_3$ critical point given by
\begin{eqnarray}
a_2&=&\frac{1}{2}\ln\left[
\frac{\sqrt{g_1(2g_2-g_1)\left(g_1^2-g_2^2+2g_1\left(g_2+\sqrt{g_2(2g_1-g_2)}\right)\right)}}
{(g_1-g_2)\left(g_1+\sqrt{g_2(2g_1-g_2)}\right)}\right.
\nonumber \\
& &\left.-\frac{g_2\left(g_1+\sqrt{g_2(2g_1-g_2)}\right)}
{(g_1-g_2)\left(g_1+\sqrt{g_2(2g_1-g_2)}\right)}\right],\nonumber \\
a_6&=&\frac{1}{2}\ln \frac{g_1+\sqrt{g_2(2g_1-g_2)}}{g_1-g_2},\qquad
V_0=\frac{64g_1^2g_2^2}{(g_1-g_2)^2}\, .
\end{eqnarray}
This critical point is invariant under $SU(2)\times SU(2)\times
U(1)^2$ symmetry.
\item There is another $dS_3$ vacuum given by $a_4=a_3=a_2$, $a_8=a_7=a_6$ and
\begin{eqnarray}
b_1&=&\ln \frac{g_2}{(g_2-g_1)\cosh ^3a_2},\qquad b_5=\ln
\frac{g_1}{(g_1-g_2)\cosh ^3a_6},
\nonumber \\
V_0&=&\frac{64g_1^2g_2^2}{(g_1-g_2)^2}
\end{eqnarray}
with $SU(2)\times SU(2)$ symmetry.
\end{itemize}

\subsection{Critical points on the $SU(2)_{L\textrm{diag}}$ invariant manifold}
We further reduce the residual symmetry to
$SU(2)_{L\textrm{diag}}\subset SU(2)^+_L\times SU(2)^-_L$. Under
$SO(4)_{\textrm{diag}}$, we already know that the 64 scalars
transform as four copies of $\mathbf{1}+\mathbf{6}+\mathbf{9}$. We
can then further truncate to $SU(2)_{L\textrm{diag}}$ and find
sixteen singlets given by four copies of
$(\mathbf{1},\mathbf{1})+(\mathbf{1},\mathbf{3})$ under
$SU(2)_{L\textrm{diag}}\times SU(2)_{R\textrm{diag}}$. They can be
parametrized by the coset representative
\begin{equation}
L=\prod_{i=1}^{16} e^{a_i\mc{Y}_i}
\end{equation}
in which the non-compact generators are defined by
\begin{eqnarray}
\mc{Y}_1&=&\frac{1}{2}\left(Y^{15}+Y^{26}+Y^{37}+Y^{48}\right),\qquad
\mc{Y}_2=\frac{1}{2}\left(Y^{16}-Y^{25}+Y^{38}-Y^{47}\right),\nonumber
\\
\mc{Y}_3&=&\frac{1}{2}\left(Y^{17}-Y^{35}-Y^{28}+Y^{46}\right),\qquad
\mc{Y}_4=\frac{1}{2}\left(Y^{18}-Y^{45}+Y^{27}-Y^{36}\right),\nonumber
\\
\mc{Y}_5&=&\frac{1}{2}\left(Y^{51}+Y^{62}+Y^{73}+Y^{84}\right),\qquad
\mc{Y}_6=\frac{1}{2}\left(Y^{52}-Y^{61}+Y^{74}-Y^{83}\right),\nonumber
\\
\mc{Y}_7&=&\frac{1}{2}\left(Y^{53}-Y^{71}-Y^{64}+Y^{82}\right),\qquad
\mc{Y}_8=\frac{1}{2}\left(Y^{54}-Y^{81}+Y^{63}-Y^{72}\right),\nonumber
\\
\mc{Y}_9&=&\frac{1}{2}\left(Y^{11}+Y^{22}+Y^{33}+Y^{44}\right),\qquad
\mc{Y}_{10}=\frac{1}{2}\left(Y^{12}-Y^{21}+Y^{34}-Y^{48}\right),\nonumber
\\
\mc{Y}_{11}&=&\frac{1}{2}\left(Y^{13}-Y^{31}-Y^{24}+Y^{42}\right),\qquad
\mc{Y}_{12}=\frac{1}{2}\left(Y^{14}-Y^{41}+Y^{23}-Y^{32}\right),\nonumber
\\
\mc{Y}_{13}&=&\frac{1}{2}\left(Y^{55}+Y^{66}+Y^{77}+Y^{88}\right),\qquad
\mc{Y}_{14}=\frac{1}{2}\left(Y^{56}-Y^{65}+Y^{78}-Y^{87}\right),\nonumber
\\
\mc{Y}_{15}&=&\frac{1}{2}\left(Y^{57}-Y^{75}-Y^{68}+Y^{86}\right),\qquad
\mc{Y}_{16}=\frac{1}{2}\left(Y^{58}-Y^{85}+Y^{67}-Y^{76}\right).\,\,\,\quad
.
\end{eqnarray}
\indent From a very complicated potential, we find one
non-supersymmetric critical point given by
\begin{eqnarray}
a_6&=&\ln
\frac{\sqrt{g_2}-\sqrt{3g_1}}{\sqrt{g_2}+\sqrt{3g_1}},\qquad
g_2=(2+\sqrt{13})g_1,\nonumber \\
V_0&=&-8(469+131\sqrt{13})g_1^2
\end{eqnarray}
which is invariant under $SU(2)\times U(1)$ symmetry.
\\
\indent Apart from $P_1$, $P_2$ and $P_3$, we have not given the
complete mass spectra for other $AdS_3$ critical points. This is
mainly because computing the full scalar masses for those critical
points is much more involved. The stability of these critical points
is uncertain without the full scalar masses. A partial check shows
that at least the scalar masses for the singlets in each sector
satisfy the BF bound. It could happen that some other scalars might
have masses violating the bound. However, similar to the three
stable critical points studied above, it is likely that the other
critical points are stable for some values of $\alpha$.

\section{Deformations of the $N=(4,4)$ SCFT}\label{flow}
In this section, we will study some RG flow solutions interpolating
between the maximally supersymmetric $SO(4)\times SO(4)$ critical
point in the UV and some of the non-supersymmetric critical points
identified in the previous section. We will also consider
supersymmetric flows to non-conformal field theories in the IR.
\subsection{Supersymmetric deformations}
We begin with supersymmetric solutions which can be obtained by
finding solutions of the associated BPS equations. We have not found
any supersymmetric critical point apart from the trivial one at
$L=\mathbf{I}$, so we only expect to find flow solutions to
non-conformal field theories. In these flows, the solutions
interpolate between the UV point at which all scalars vanish and the
IR with infinite values of scalar vev's \cite{non_CFT_flow}. Since
supersymmetric solutions are of interest here, we need the
supersymmetry transformations of fermions which in the present case
are given by the non-propagating gravitini $\psi^{I}_\mu$ and the
spin-$\frac{1}{2}$ fields $\chi^{iI}$. Their supersymmetry
transformations are given by, see \cite{dewit} for more details and
conventions,
\begin{eqnarray}
\delta\psi^I_\mu
&=&\mathcal{D}_\mu\epsilon^I+gA_1^{IJ}\gamma_\mu\epsilon^J,\label{d_chi}\\
\delta\chi^{iI}&=&
\frac{1}{2}(\delta^{IJ}\mathbf{1}-f^{IJ})^i_{\phantom{a}j}{\mathcal{D}{\!\!\!\!/}}\phi^j\epsilon^J
-gNA_2^{JIi}\epsilon^J\label{d_psi}\, .
\end{eqnarray}
These equations will be used to find supersymmetric solutions in the
next subsections.
\subsubsection{A supersymmetric flow to $SO(4)\times SO(4)$ non-conformal field theory}
We first look for a simple solution preserving $SO(4)\times SO(4)$
symmetry. Accordingly, only $a_1$ and $a_2$ in equation
\eqref{SO4d_coset} are turned on in order to preserve the full
$SO(4)\times SO(4)$. Using the standard domain wall ansatz for the
metric
\begin{equation}
ds^2=e^{2A}dx^2_{1,1}+dr^2\label{3D_metric_ansatz}
\end{equation}
with $A$ depending only on the radial coordinate $r$, we find the
BPS equations
\begin{eqnarray}
a_1'+8g_1e^{2a_1}\left(e^{2a_1}-1\right)&=&0,\label{a1p}\\
a_2'+8g_2e^{2a_2}\left(e^{2a_2}-1\right)&=&0,\label{a2p}\\
A'+8\left[g_1e^{2a_1}\left(e^{2a_1}-2\right)+g_2e^{2a_2}\left(e^{2a_2}-2\right)\right]&=&0\label{Ap}
\end{eqnarray}
where we have imposed the projector $\gamma_r
\epsilon^I=-\epsilon^I$, $I=2,4,5,8$ and $\gamma_r
\epsilon^I=\epsilon^I$, $I=1,3,6,7$. The $'$ denotes the
$r$-derivative. The resulting solution is then half-supersymmetric
with $N=(4,4)$ Poincare supersymmetry in the dual two dimensional
field theory. Equations \eqref{a1p} and \eqref{a2p} can be solved
for $a_1$ and $a_2$ as an implicit function of $r$. The result is
\begin{eqnarray}
r&=&c_1-\frac{1}{16g_1}\left[e^{-2a_1}+\ln \left(1-e^{-2a_1}\right)\right],\label{a1_sol}\\
r&=&c_2-\frac{1}{16g_2}\left[e^{-2a_2}+\ln
\left(1-e^{-2a_2}\right)\right]\label{a2_sol}
\end{eqnarray}
with integration constants $c_1$ and $c_2$. Equation \eqref{Ap} can
immediately be integrated to give $A$ as a function of $a_1$ and
$a_2$. The result is
\begin{equation}
A=2(a_1+a_2)-\frac{1}{2}\ln (1-e^{2a_1})-\frac{1}{2}\ln
(1-e^{2a_2})\, .\label{A_sol}
\end{equation}
\indent In the UV, the dual field theory is conformal with
$a_1=a_2=0$. Near this point, the scalars behave as $a_1\approx
e^{-16g_1r}=e^{-\frac{2g_1}{g_1+g_2}\frac{r}{L_{UV}}}$ and
$a_2\approx e^{-16g_2r}=e^{-\frac{2g_2}{g_1+g_2}\frac{r}{L_{UV}}}$.
We see that $a_{1,2}\rightarrow 0$ as $r\rightarrow \infty$. In this
limit, we find $A'\approx8(g_1+g_2)=\frac{1}{L_{UV}}$ or $A\approx
\frac{r}{L_{UV}}$ which gives the maximally supersymmetric $AdS_3$.
\\
\indent As $a_1,a_2\rightarrow \infty$, we find $r\rightarrow
\textrm{constant}$ as it should. Near $a_1,a_2\rightarrow \infty$,
equations \eqref{a1_sol} and \eqref{a2_sol} give $a_1\approx
-\frac{1}{4}\ln \left(32g_1r\right)$ and $a_2\approx -\frac{1}{4}\ln
\left(32g_2r\right)$. From equation \eqref{A_sol}, we find $A\approx
a_1+a_2=-\frac{1}{4}\ln \left[(32r)^2g_1g_2\right]$. Accordingly,
the metric becomes a domain wall in the IR
\begin{equation}
ds^2=\frac{1}{32r\sqrt{g_1g_2}}dx^2_{1,1}+dr^2\, .
\end{equation}
The full bosonic symmetry is $ISO(1,1)\times SO(4)\times SO(4)$
corresponding to non-comformal field theory with $N=(4,4)$
supersymmetry.
\\
\indent However, flows of this type generally involve singularities.
Various types of possible singularities have been classified in
\cite{Gubser_singularity}. According to the result of
\cite{Gubser_singularity}, physical singularities are the ones at
which the scalar potential is bounded from above. However, with the
solution given above, the potential becomes infinite in this case.
Therefore, the corresponding flow solution is not physically
acceptable by the criterion of \cite{Gubser_singularity}. Since the
framework we have used could be uplifted to ten dimensions via
$S^3\times S^3\times S^1$ reduction, it is interesting to
investigate whether this singularity is resolved in the full string
theory.

\subsubsection{A half-supersymmetric domain wall}
We then look for a more general supersymmetric solution. The scalar
sector of interest here is the $SU(2)^+_L\times SU(2)^-_L$ invariant
one given in \eqref{L_SU2_SU2}. We first relabel the scalars
$(a_2,a_3,a_4,a_6,a_7,a_8)$ to $(b_2,b_3,b_4,b_6,b_7,b_8)$ in order
to work with a uniform notation.
\\
\indent We begin with the BPS equations given by $\delta\chi^{iI}=0$
\begin{eqnarray}
b_1'&=&-16g_1e^{b_1}\left(e^{b_1}-\textrm{sech}b_2 \textrm{sech}b_3\textrm{sech}b_4\right),\label{eq_b1}\\
b_2'&=&-16g_1e^{b_1}\left(e^{b_1}\cosh b_2-\textrm{sech}b_3\textrm{sech}b_4\right)\sinh b_2,\label{eq_b2}\\
b_3'&=&-16g_1\cosh b_2\sinh b_3e^{b_1}\left(e^{b_1}\cosh b_2\cosh b_3-\textrm{sech}b_4\right),\label{eq_b3}\\
b_4'&=&-16g_1\cosh b_2\cosh b_3\sinh b_4 e^{b_1}\left(e^{b_1}\cosh b_2 \cosh b_3 \cosh b_4-1\right),\label{eq_b4}\\
b_5' &=&-16g_2e^{b_5}\left(e^{b_5}-\textrm{sech}b_6\textrm{sech}b_7\textrm{sech}b_8\right),\label{eq_b5}\\
b_6'&=&-16g_2\sinh b_6 e^{b_5}\left(e^{b_5}\cosh b_6-\textrm{sech}b_7\textrm{sech}b_8\right),\label{eq_b6}\\
b_7'&=&-16g_2\cosh b_6\sinh b_7e^{b_5}\left(e^{b_5}\cosh b_6\cosh b_7-\textrm{sech}b_8\right),\label{eq_b7}\\
b_8'&=&-16g_2\cosh b_6\cosh b_7\sinh b_8e^{b_5}\left(e^{b_5}\cosh
b_6\cosh b_7\cosh b_8-1\right).\label{eq_b8}
\end{eqnarray}
where we have used the projection conditions $\gamma_r
\epsilon^I=-\epsilon^I$, $I=2,4,5,8$ and $\gamma_r
\epsilon^I=\epsilon^I$, $I=1,3,6,7$ as in the previous case. The
gravitino variation $\delta \psi^I_\mu$, $\mu=0,1$, gives
\begin{eqnarray}
A'&=&-8g_1e^{b_1}\cosh b_2\cosh b_3\cosh b_4\left(e^{b_1}\cosh b_2\cosh b_3\cosh b_4-2\right)\nonumber \\
& &-8g_2e^{b_5}\cosh b_6\cosh b_7\cosh b_8 \left(e^{b_5}\cosh
b_6\cosh b_7\cosh b_8-2\right).\label{eq_A}
\end{eqnarray}
\indent From these equations, we see that apart from the maximally
supersymmetric point at $b_i=0$, $i=1,\ldots, 8$, there is a flat
direction of the potential given by
\begin{equation}
e^{-b_1}=\cosh b_2\cosh b_3\cosh b_4,\qquad e^{-b_5}=\cosh b_6\cosh
b_7\cosh b_8
\end{equation}
which leads to $V_0=-64(g_1+g_2)^2$. Equation \eqref{eq_A} gives
$A'=8(g_1+g_2)$ or $A=8(g_1+g_2)r$ which is the $AdS_3$ solution
with radius $L=\frac{1}{8(g_1+g_2)}$. It can also be verified that
the full $(4,4)$ supersymmetry is preserved. This should correspond
to a marginal deformation of the $N=(4,4)$ SCFT. There are no other
supersymmetric critical points in this sector. Therefore, the flow
breaking supersymmetry from $(4,4)$ to $(4,0)$ is not possible.
\\
\indent However, there is a half-supersymmetric domain wall solution
similar to the dilatonic p-brane solutions of $N=1$, $D=7$ and
$N=2$, $D=6$ gauged supergravities studied in
\cite{dilatonic_p_brane}. It is remarkable that the full set of the
above equations admits an analytic solution. The strategy to find
the solution is as follow. We first determine $b_{2,3,4}$ as
functions of $b_1$ and similarly determine $b_{6,7,8}$ as functions
of $b_5$. $b_1$ and $b_5$ are determined as functions of $r$ and can
be solved explicitly. From \eqref{eq_b1} and \eqref{eq_b2}, we find
\begin{equation}
\frac{db_2}{db_1}=\cosh b_2\sinh b_2
\end{equation}
which can be solved for $b_2$ as a function of $b_1$ giving rise to
\begin{equation}
b_2=\coth^{-1}e^{-b_2-2c_1}\, .\label{b2_sol}
\end{equation}
Using \eqref{eq_b1} and \eqref{eq_b3} together with $b_2$ solution
from \eqref{b2_sol}, we find
\begin{equation}
\frac{db_3}{db_1}=\frac{\sinh (2b_3)}{2\left(1-e^{2b_1+4c_1}\right)}
\end{equation}
whose solution is given by
\begin{equation}
b_3=\tanh^{-1}\frac{e^{b_1+2c_2}}{\sqrt{1-e^{2b_1+4c_1}}}\,
.\label{b3_sol}
\end{equation}
Combining \eqref{eq_b1} and \eqref{eq_b4} and substituting for $b_2$
and $b_3$ solutions give
\begin{equation}
\frac{db_4}{db_1}=-\frac{\cosh b_4 \sinh
b_4}{\left(e^{4c_1}+e^{4c_2}\right)e^{b_1}-1}\, .
\end{equation}
We then find the solution for $b_4$
\begin{equation}
b_4=\tanh^{-1}\frac{e^{b_1+2c_3}}{\sqrt{1-e^{2b_1}\left(e^{4c_1}+e^{4c_2}\right)}}\,
.\label{b4_sol}
\end{equation}
With solutions for $b_2$, $b_3$ and $b_4$, equation \eqref{eq_b1}
becomes
\begin{equation}
b_1'=16g_1e^{b_1}\left(\sqrt{1-e^{2b_1}\left(e^{4c_1}+e^{4c_2}+e^{4c_3}\right)}-e^{b_1}\right).
\end{equation}
This can be solved for $b_1$ as an implicit function of $r$. The
solution is
\begin{eqnarray}
r&=&-\frac{1}{32g_1}\left[2e^{-b_1}\sqrt{1-\beta_1 e^{2b_1}}+\ln
\left[e^{-2b_1}\left((\beta_1-1)e^{2b_1}-1+2e^{b_1}
\sqrt{1-\beta_1}e^{2b_1}\right)\right]\right]\nonumber \\
& &+\textrm{constant}
\end{eqnarray}
where $\beta_1=e^{4c_1}+e^{4c_2}+e^{4c_3}$.
\\
\indent We can solve \eqref{eq_b5} to \eqref{eq_b8} by the same
procedure. The resulting solutions are given by
\begin{eqnarray}
b_6 &=&\tanh^{-1} e^{b_5+2c_4},\qquad b_7=\tanh^{-1}\frac{e^{b_5+2c_5}}{\sqrt{1-e^{2b_5+4c_4}}},\nonumber \\
b_8&=& \tanh^{-1}\frac{e^{b_5+3c_6}}{\sqrt{1-e^{b_5}\left(e^{4c_4}+e^{4c_5}\right)}},\nonumber \\
r&=&-\frac{1}{32g_2}\left[2e^{-b_5}\sqrt{1-\beta_2e^{2b_5}}+\ln
\left[e^{-2b_5}\left((\beta_2-1)e^{2b_5}-1+
2e^{b_5}\sqrt{1-\beta_2e^{2b_5}}\right)\right]\right]\nonumber \\
& &+\textrm{constant}
\end{eqnarray}
where $\beta_2=e^{4c_4}+e^{4c_5}+e^{4c_6}$.
\\
\indent After substituting all of the $b_i$ solutions for
$i=2,3,4,6,7,8$ in \eqref{eq_A}, we obtain
\begin{equation}
A'=\frac{16g_1e^{b_1}}{\sqrt{1-\beta_1
e^{2b_1}}}-\frac{8g_1e^{2b_1}}{1-\beta_1 e^{2b_1}}
+\frac{16g_2e^{b_5}}{\sqrt{1-\beta_2
e^{2b_5}}}-\frac{8g_2e^{2b_5}}{1-\beta_2 e^{2b_5}}
\end{equation}
whose solution in terms of $b_1$ and $b_5$ is readily found by a
direct integration using \eqref{eq_b1} and \eqref{eq_b5} including
the solutions for the other $b_i$'s. The resulting solution is given
by
\begin{eqnarray}
A&=&b_1+b_5+\frac{1}{2}\tanh^{-1}\frac{e^{b_1}}{\sqrt{1-\beta_1e^{2b_1}}}
+\frac{1}{2}\tanh^{-1}\frac{e^{b_5}}{\sqrt{1-\beta_2e^{2b_5}}}-\ln \left[1-\beta_1e^{2b_1}\right]\nonumber \\
& &-\ln \left[1-(1+\beta_1)e^{2b_1}\right] -\ln
\left[1-\beta_2e^{2b_5}\right]-\ln
\left[1-(1+\beta_2)e^{2b_5}\right].
\end{eqnarray}
\indent As $b_1,b_5\rightarrow 0$, other scalars do not vanish for
finite $c_i$. We then find that the solution will not have an
interpretation in terms of the usual holographic RG flows. The
solution is rather of the 1-brane soliton type, see
\cite{dilatonic_p_brane} for a general discussion of $(D-2)$-brane
solitons in $D$ dimensions. It can also be verified that the $\delta
\psi^I_r=0$ condition precisely gives the Killing spinors for the
unbroken supersymmetry $\epsilon^I=e^{\frac{A}{2}}\epsilon^I_0$ with
the constant spinor $\epsilon^I_0$ satisfying
$\gamma_r\epsilon_0^I=-\epsilon_0^I$, $I=2,4,5,8$ and
$\gamma_r\epsilon_0^I=\epsilon_0^I$, $I=1,3,6,7$.

\subsection{Non-supersymmetric deformations}
We now study non-supersymmetric RG flow solutions interpolating
between the $N=(4,4)$ SCFT in the UV and some critical points found
in the previous section. The solutions are essentially
non-supersymmetric since they connect a supersymmetric to a
non-supersymmetric critical point. Finding the corresponding
solutions involve solving the full second order field equations for
both the scalars and the metric in contrast to solving the first
order BPS equations in the supersymmetric case. Although there are
some examples of analytic supersymmetric flow solutions in three
dimensions, in general, analytic solutions with many active scalars,
even for the supersymmetric case, can be very difficult to find.
Therefore, we will not expect to find any analytic solutions in the
non-supersymmetric case but rather look for numerical flow
solutions.
\\
\indent We begin with the scalar and metric Lagrangian which in our
convention is given by a truncation of the bosonic part of the
gauged Lagrangian given in \cite{dewit}
\begin{equation}
\mc{L}=\frac{1}{2}R-\frac{1}{2}g_{ij}\pd_\mu \phi^i \pd^\mu
\phi^j-V\, .
\end{equation}
In the rest of this section, we will work with the canonically
normalized kinetic terms for scalars. With the standard domain wall
ansatz for the metric in \eqref{3D_metric_ansatz}, the field
equations for $d$ scalars $\phi^i(r)$ and the metric function $A(r)$
can be found to be
\begin{eqnarray}
{\phi^i}''+2A'{\phi^i}'-\frac{\pd V}{\pd \phi^i}&=&0, \label{phi_i_eq}\\
2A'^2-\sum_{i=1}^d({\phi^i}')^2+2V &=&0,\label{Ap_eq}\\
2A''+2A'^2+\sum_{i=1}^d({\phi^i}')^2+2V&=&0\label{App_eq}\, .
\end{eqnarray}
Note that equations \eqref{Ap_eq} and \eqref{App_eq} can be combined
to $A''=-\sum_{i=1}^d({\phi^i}')^2$. There are indeed only $d+1$
independent equations since \eqref{App_eq} can be obtained by
differentiating \eqref{Ap_eq} and using \eqref{phi_i_eq}.
\\
\indent Our flows mainly involve one active scalar. We will then
concentrate on this case. The general solution to the second order
field equation near the UV point is of the form
\begin{equation}
\phi=Ae^{-\frac{(d-\Delta)r}{L}}+Be^{-\frac{\Delta r}{L}}
\end{equation}
for which $d=2$ in the present case. The first and second terms
correspond to turning on an operator of dimension $\Delta$ and a
vacuum expectation value (vev.) of the operator, respectively. In
contrast to supersymmetric flows obtained by solving first order BPS
equations which result in one of the two possibilities, there is
some ambiguity in non-supersymmetric flows. Both of the two terms in
the above equation arise in the behavior of $\phi$ near the UV
point.
\\
\indent One way to solve this ambiguity is to recast the second
order field equations into a first order form by introducing the
generating function $W$ \cite{de_Boer_RGflow}, see also a review
\cite{de_Boer_RGnote},
\begin{eqnarray}
\phi'&=&-\frac{\pd W}{\pd \phi},\label{first_eq1}\\
A'&=&W\label{first_eq2}\, .
\end{eqnarray}
$W$ is related to the scalar potential by the relation
\begin{equation}
V=\frac{1}{2}\left(\frac{\pd W}{\pd\phi}\right)^2-W^2\, .
\end{equation}
With the generating function $W$ defined above, it is not difficult
to show that \eqref{first_eq1} and \eqref{first_eq2} with an obvious
generalization to $d$ scalars lead to \eqref{phi_i_eq} and
\eqref{Ap_eq}.
\\
\indent In supersymmetric cases, $W$ becomes the true superpotential
related to the $A_1^{IJ}$ tensor in our case, and flow solutions
interpolate between critical points of $W$. For non-supersymmetric
flows, the flows interpolate between some critical points of $V$
which are not critical points of the superpotential. However, after
the flow solutions (usually the numerical ones) are found, the
corresponding generating function is obtained by solving
\eqref{first_eq1}. Given the generating function, the first order
equation \eqref{first_eq1} can be used to determine the precise
behavior near the UV point and eventually leads to the resolution of
operator and vev deformations.
\\
\indent From \cite{CFT_CFT_flow}, the generating function has an
expansion near the UV point as
\begin{equation}
W=-\frac{2(d-1)}{L}+\frac{(d-\Delta)}{2L}\phi^2+\ldots
\end{equation}
for operator deformations and
\begin{equation}
W=-\frac{2(d-1)}{L}+\frac{\Delta}{2L}\phi^2+\ldots
\end{equation}
for vev deformations. Using \eqref{first_eq1}, we find that near the
UV point, normally at $\phi=0$,
\begin{equation}
\frac{\phi'}{\phi}\approx \left\{\begin{array}{cc}
               -\frac{d-\Delta}{L},&\qquad \textrm{for operator deformations} \\
               -\frac{\Delta}{L},& \qquad \textrm{for vev deformations}
             \end{array}\right.\, .\label{UV_behavior}
\end{equation}
Therefore, we can determine whether our flows are driven by turning
on an operator or by a vev of an operator by using
\eqref{UV_behavior}. A similar study of non-supersymmetric RG flows
in $N=2$ three dimensional gauged supergravity has also been done in
\cite{deger}.

\subsubsection{Flow to $SO(4)\times SO(4)$ CFT}
\begin{figure}[!h] \centering
\includegraphics[width=0.5\textwidth, bb = 0 0 220 170 ]{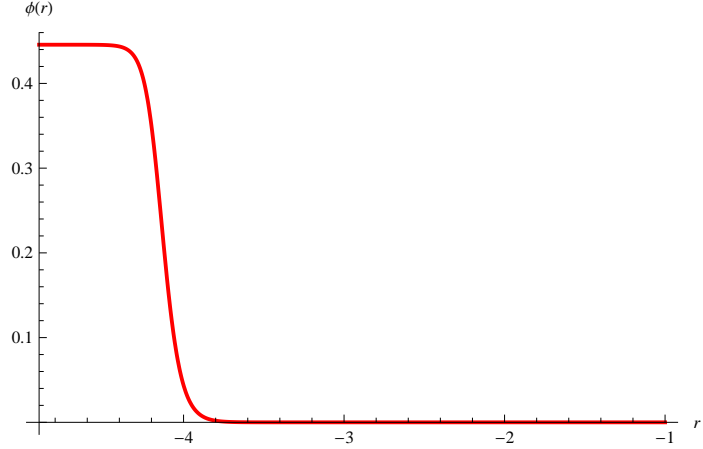}\\
\caption{A solution for $\phi$ in a flow to $SO(4)\times SO(4)$ CFT
with $g_1=1$ and $g_2=-4$.} \label{fig1}
\end{figure}

\begin{figure}[!h] \centering
\includegraphics[width=0.5\textwidth, bb = 0 0 220 170 ]{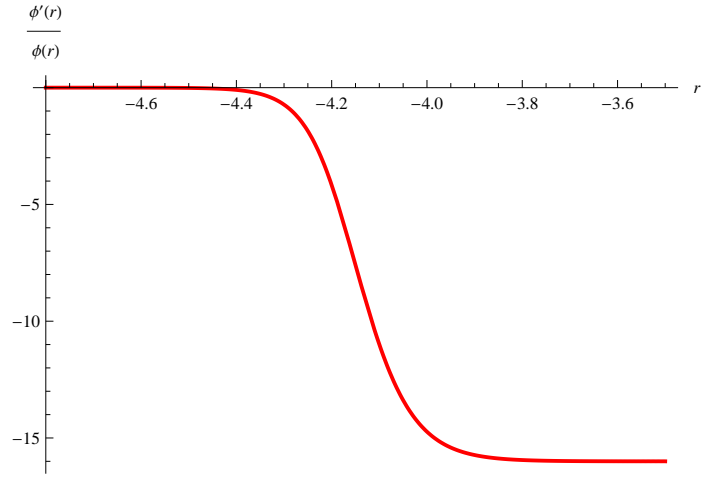}\\
\caption{The behavior of $\frac{\phi'}{\phi}$ in the flow to
$SO(4)\times SO(4)$ CFT with $g_1=1$ and $ g_2=-4$.} \label{fig2}
\end{figure}

The IR fixed point of this flow is critical point $P_1$
corresponding to a CFT with $SO(4)\times SO(4)$ symmetry. The field equations can be consistently truncated to a
single scalar $\phi=a_1$. We find the field equations
\begin{eqnarray}
\phi''+2\phi'A'-64\left(4g_1^2e^{4\phi}-4g_1(g_1-g_2)e^{2\phi}-4g_1g_2e^{\phi}\right)&=&0,\\
2A'^2-\phi'^2+128\left(g_1^2e^{4\phi}-2g_1(g_1-g_2)e^{2\phi}-4g_1g_2e^{\phi}-g_2^2\right)&=&0\,
.
\end{eqnarray}
The flow solution will interpolate between $\phi_{UV}=0$ in the UV
and $\phi_{IR}=\ln \frac{\sqrt{g_1(g_1-4g_2)}-g_1}{2g_1}$ in the IR.
A numerical solution to the above equations can be found, and an
example solution for $g_1=1$ and $g_2=-4$ is shown in Figure
\ref{fig1}. It can be explicitly seen that the flow interpolates
between $\phi_{UV}=0$ and $\phi_{IR}=0.445681$ at this value of
$g_1$ and $g_2$. Figure \ref{fig2} shows the behavior of
$\frac{\phi'}{\phi}$ along the flow. With $g_1=1$ and $g_2=-4$, the
scalar mass gives the dimension of the dual operator
$\Delta=\frac{4}{3}$. The $AdS_3$ radius in the UV is
$L_{UV}=\frac{1}{8|g_1+g_2|}=\frac{1}{24}$. From Figure \ref{fig2},
we find that, near the UV point, $\frac{\phi'}{\phi}=-15.984...=
-\frac{0.666...}{L_{UV}}$ or $\phi\sim e^{-\frac{2r}{3L_{UV}}}$.
Therefore, the flow is driven by a relevant operator of dimensions
$\frac{4}{3}$ and corresponds to a true deformation rather than a
deformation by a vacuum expectation value. With these values of the
coupling constants, the ratio of the central charges is
$\frac{c_{UV}}{c_{IR}}=1.025$.
\\

\subsubsection{Flow to $SO(4)$ CFT} 
\begin{figure}[!h] \centering
\includegraphics[width=0.5\textwidth, bb = 0 0 220 170 ]{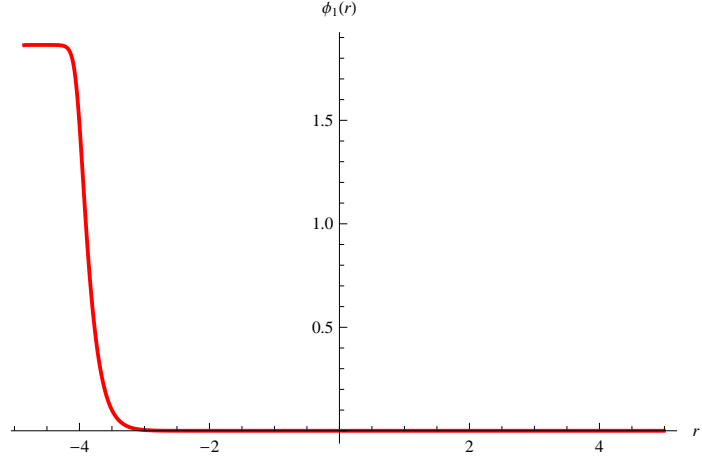}\\
\caption{A solution for $\phi_1$ in a flow to $SO(4)$ CFT with
$g_1=1$.} \label{fig3}
\end{figure}

\begin{figure}[!h] \centering
\includegraphics[width=0.5\textwidth, bb = 0 0 220 170 ]{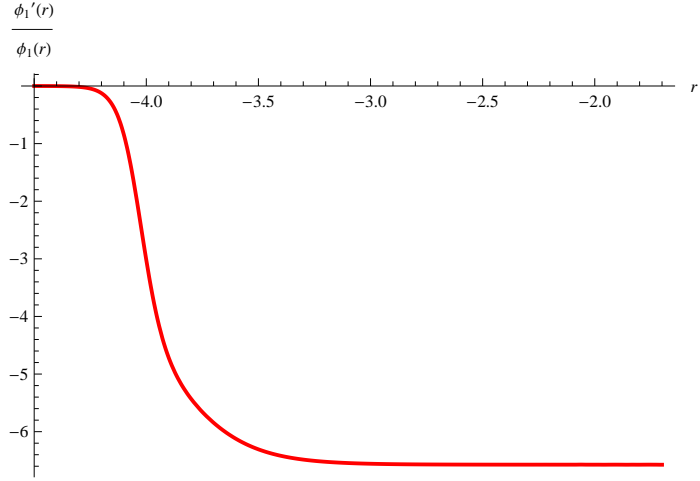}\\
\caption{The behavior of $\frac{\phi'_1}{\phi_1}$ in the flow to
$SO(4)$ CFT with $g_1=1$.} \label{fig4}
\end{figure}

We now consider an RG flow to
critical point $P_2$ with residual gauge symmetry
$SO(4)_\textrm{diag}$. In this case, the parameter $\alpha$ is
positive, and the flow can be regarded as a flow to another
$AdS_3\times S^3\times S^3$ background in the IR. The field
equations are again consistently truncated to a single scalar which
we will call $\phi_1$. Using $g_2=\frac{\sqrt{13}-2}{9}g_1$ and
$g_1=1$, we find
\begin{equation}
\phi_{1IR}=1.8641,\qquad L_{UV}=\frac{7-\sqrt{13}}{32}\approx
0.1061,\qquad \frac{c_{UV}}{c_{IR}}\approx 1.6422\, .
\end{equation}
That the flow is driven by this
particular $SO(4)$ singlet among all of the four singlets can be
seen by looking at the masses of the four singlets at the UV point.
With the above values of $g_1$ and $g_2$ or more precisely the value
of $\alpha$, only the singlet associated to our critical point is
tachyonic corresponding to a relevant dual operator of dimension
$\Delta=1.303$.
\\
\indent The field equations are given by
\begin{eqnarray}
\phi_1''+2\phi_1'A'+8\cosh^4\frac{\phi_1}{2}\left[4(g_1^2-6g_1g_2+g_2^2)\sinh\phi_1-2(g_1-3g_2)^2\sinh(2\phi_1)\right]& &\nonumber \\
+16\sinh\frac{\phi_1}{2}\cosh^3\frac{\phi_1}{2}\left[5g_1^2+34g_1g_2+13g_2^2+4(g_1^2-6g_1g_2+g_2^2)\cosh\phi_1\right.& &\nonumber \\
\left.-(g_1-3g_2)^2\cosh(2\phi_1)\right]=0,& &\\
2A'^2-\phi_1'^2-16\cosh^4\frac{\phi_1}{2}\left[5g_1^2+34g_1g_2+13g_2^2+4(g_1^2-6g_1g_2+g_2^2)\cosh\phi_1\right.& &\nonumber \\
\left.-(g_1-3g_2)^2\cosh(2\phi_1)\right]=0\, .& &
\end{eqnarray}
A flow solution with $g_1=1$ is shown in Figure \ref{fig3}.
\\
\indent Near the UV point, we find that
$\frac{\phi_1'}{\phi_1}=-6.5729=-\frac{0.697}{L_{UV}}$ as shown in
Figure \ref{fig4}. The flow is then driven by a relevant operator of
dimension $1.303$ since $\phi_1\sim e^{-\frac{0.697r}{L_{UV}}}$ near
the UV point. As in the previous solution, the flow describes a true
deformation of the UV SCFT. \vspace{1cm}

\subsubsection{Flow to $SO(4)\times SO(2)\times SO(2)$ CFT}
As in the previous subsections, we begin with the scalar field
equations. It is easily verified that setting all but
$(a_1,a_2,a_3,a_4)$ to zero satisfies their field equations. The
field equations of these scalars and the metric are given by
\begin{eqnarray}
a_1''+2a_1'A'
-128\sqrt{2}g_1\left[g_1e^{\sqrt{2}(a_1+a_2)}\left(e^{\sqrt{2}(a_1+a_2)}-1\right)\right.\qquad \qquad& &\nonumber \\
\left.g_2e^{\sqrt{2}a_1}\left(e^{\sqrt{2}a_2}-1\right)\left(e^{\sqrt{2}a_3}+e^{\sqrt{2}a_4}-e^{\sqrt{2}(a_3+a_4)}\right)\right]&=&
0,\\
a_2''+2a_2'A'-128\sqrt{2}g_1\left[g_1e^{\sqrt{2}a_1+a_2}\left(e^{\sqrt{2}(a_1+a_2)}-1\right)\right.\qquad \qquad & &\nonumber \\
\left.
+g_2e^{\sqrt{2}a_2}\left(e^{\sqrt{2}a_1}-1\right)\left(e^{\sqrt{2}a_3}+e^{\sqrt{2}a_4}-
e^{\sqrt{2}(a_3+a_4)}\right)\right]&=&0,\\
a_3''+2a_3'A'+128\sqrt{2}g_2\left[g_2\left(e^{\sqrt{2}(a_3+a_4)}-e^{2\sqrt{2}(a_3+a_4)}\right)\right.\qquad \qquad & &\nonumber \\
\left.
+g_1e^{\sqrt{2}a_3}\left(e^{\sqrt{2}a_4}-1\right)\left(e^{\sqrt{2}a_1}+e^{\sqrt{2}a_2}
-e^{\sqrt{2}(a_1+a_2)}\right) \right]&=&0,\\
a_4''+2a_4'A'+128\sqrt{2}g_2\left[g_2\left(e^{\sqrt{2}(a_3+a_4)}-e^{2\sqrt{2}(a_3+a_4)}\right)\right.\qquad \qquad & &\nonumber \\
\left.-g_1e^{\sqrt{2}a_4}\left(e^{\sqrt{2}a_3}-1\right)\left(e^{\sqrt{2}a_1}+e^{\sqrt{2}a_2}
-e^{\sqrt{2}(a_1+a_2)}\right) \right]&=&0,\\
2A'^2-a_1'^2-a_2'^2-a_3'^2-a_4'^2\qquad \qquad \qquad \qquad \qquad \qquad \qquad \qquad & &\nonumber \\
+128\left[g_1^2\left(e^{2\sqrt{2}(a_1+a_2)}-2e^{\sqrt{2}(a_1+a_2)}\right)+g_2^2\left(e^{2\sqrt{2}(a_2+a_3)}-2
e^{\sqrt{2}(a_2+a_3)}\right)\right.& &\nonumber \\
\left.-256g_1g_2\left(e^{\sqrt{2}a_1}+4e^{\sqrt{2}a_2}-e^{\sqrt{2}(a_1+a_2)}\right)
\left(e^{\sqrt{2}a_3}+e^{\sqrt{2}a_4}-e^{\sqrt{2}(a_3+a_4)}\right)\right]
&=&0\,. \qquad
\end{eqnarray}
It can be checked that a truncation $a_2=a_1=a$ is also consistent,
but $a_4=-a_3$ is not. So, together with the equation for $A$, we
are effectively left with four equations to be solved. However,
these four equations are still complicated. We will not attempt to
give their solution here but simply end this section with some
comments on the flow.
\\
\indent In terms of the $SO(4)\times SO(4)$ representations in the
UV, $a_3$ and $a_4$ are combinations of $(\mathbf{1},\mathbf{1})$
and $(\mathbf{1},\mathbf{9})$. From the value of $g_1$ and $g_2$ in
the stability range, it can be checked that only the deformation
dual to $a$ is relevant. The flow should mainly be driven by the
operator dual to $a$. The deformations corresponding to $a_3$ and
$a_4$ are given by vacuum expectation values of irrelevant operators
since $a_3$ and $a_4$ have positive mass squares.
\section{Conclusions and discussions}\label{conclusion}
In this paper, we have studied $N=8$ gauged supergravity in three
dimensions with a non-semisimple gauge group $(SO(4)\times
SO(4))\ltimes \mathbf{T}^{12}$. The ratio of the coupling constants
of the two $SO(4)$'s is given by a parameter $\alpha$. For positive
$\alpha$, the theory describes an effective theory of ten
dimensional supergravity reduced on $S^3\times S^3\times S^1$. For
negative $\alpha$, on the other hand, the theory may describe a
similar reduction on $S^3\times H^3\times S^1$ in which $H^3$ is a
three-dimensional hyperbolic space. With $\alpha=-1$, the
cosmological constant is zero. This solution should describe a ten
dimensional background $M_3\times S^3\times H^3\times S^1$ where
$M_3$ is the three dimensional Minkowski space.
\\
\indent We have studied the scalar potential and found a number of
non-supersymmetric critical points. The trivial critical point with
maximal supersymmetry is identified with the dual large $N=(4,4)$
SCFT in two dimensions. We have explicitly checked the stability of
some non-supersymmetric critical points by computing the full scalar
mass spectra at the critical points. They are perturbatively stable
for some values of $\alpha$ parameter in the sense that all scalar
masses are above the BF bound. It is also interesting to see whether
other critical points are stable or not. The RG flows interpolating
between the large $N=(4,4)$ SCFT in the UV and non-supersymmetric IR
fixed points have been given for the flows to $SO(4)\times SO(4)$
and $SO(4)$ CFT's. We have also resolved the ambiguity between the
operator and vev deformations arising from solving the second order
field equations and found that the flows are driven by turning on
relevant operators.
\\
\indent Another result of this paper is half-supersymmetric domain
wall solutions to $N=8$ gauged supergravity. For the domain wall
preserving $SO(4)\times SO(4)$ symmetry, the solution describes an
RG flow from $N=(4,4)$ SCFT in the UV to a non-conformal $N=(4,4)$
field theory in the IR. The solution has however a bad singularity
according to the criterion of \cite{Gubser_singularity}. For the
solution preserving $SU(2)\times SU(2)$ symmetry, the holographic
interpretation is not clear. In the point of view of a $(D-2)$-brane
soliton, the solution should describe a 1-brane soliton in three
dimensions according to the general discussion in
\cite{dilatonic_p_brane}. When uplifted to ten dimensions, the
solution might describe some configuration of D1-branes. Hopefully,
the solutions obtained in this paper might be useful in string/M
theory context, black hole physics and the AdS/CFT correspondence.
The uplifted solution of the non-conformal flow preserving
$SO(4)\times SO(4)$ symmetry is also necessary for the resolution of
its singularity if the full ten-dimensional solution turns out to be
non-singular.
\\
\indent Finally, the chiral supersymmetry breaking $(4,4)\rightarrow
(4,0)$ found in \cite{Morales} cannot be implemented in the
framework of $N=8$ gauged supergravity studied here. It would
probably require a larger theory of $N=16$ gauged supergravity with
$(SO(4)\times SO(4))\ltimes (\mathbf{T}^{12},\hat{\mathbf{T}}^{34})
$ gauge group studied in \cite{Hohm_henning}. It would be very
interesting to find the flow solution of \cite{Morales} explicitly
in the three dimensional framework. We hope to come back to these
issues in future research.
\acknowledgments The author would like to thank Henning Samtleben
for valuable correspondence. This work is partially supported by
Thailand Center of Excellence in Physics through the
ThEP/CU/2-RE3/12 project, Chulalongkorn University through
Ratchadapisek Sompote Endowment Fund under grant GDNS57-003-23-002
and The Thailand Research Fund (TRF) under grant TRG5680010.

\appendix
\section{Useful formulae and details}\label{detail}
For completeness, we include a short review of gauged supergravity
in three dimensions in the formulation of \cite{dewit}. The theory
is a gauged version of a supersymmetric non-linear sigma model
coupled to non-propagating supergravity fields. N-extended
supersymmetry requires the presence of $N-1$ almost complex
structures $f^P$, $P=2,\ldots, N$ on the scalar manifold. The
tensors $f^{IJ}=f^{[IJ]}$, generating the $SO(N)$ R-symmetry in a
spinor representation under which scalar fields transform, play an
important role. In the case of symmetric scalar manifolds of the
form $G/SO(N)\times H'$, they can be written in terms of $SO(N)$
gamma matrices. In our case, we use the $16\times 16$ Dirac gamma
matrices of $SO(8)$
\begin{equation}
\gamma^I=\left(
           \begin{array}{cc}
             0 & \Gamma^I \\
             (\Gamma^I)^T & 0 \\
           \end{array}
         \right).
\end{equation}
The $8\times 8$ gamma matrices are explicitly given by
\begin{eqnarray}
\Gamma_1 &=&\sigma_4\otimes \sigma_4\otimes \sigma_4,\qquad \Gamma_2
=\sigma_1\otimes \sigma_3\otimes \sigma_4,\nonumber \\
\Gamma_3 &=&\sigma_4\otimes \sigma_1\otimes \sigma_3,\qquad \Gamma_4
=\sigma_3\otimes \sigma_4\otimes \sigma_1,\nonumber \\
\Gamma_5 &=&\sigma_1\otimes \sigma_2\otimes \sigma_4,\qquad \Gamma_6
=\sigma_4\otimes \sigma_1\otimes \sigma_2,\nonumber \\
\Gamma_7 &=&\sigma_2\otimes \sigma_4\otimes \sigma_1,\qquad \Gamma_8
=\sigma_1\otimes \sigma_1\otimes \sigma_1
\end{eqnarray}
where
\begin{eqnarray}
\sigma_1&=&\left(
             \begin{array}{cc}
               1 & 0 \\
               0 & 1 \\
             \end{array}
           \right),\qquad
\sigma_2=\left(
             \begin{array}{cc}
               0 & 1 \\
               1 & 0 \\
             \end{array}
           \right),\nonumber \\
\sigma_3&=&\left(
             \begin{array}{cc}
               1 & 0 \\
               0 & -1 \\
             \end{array}
           \right),\qquad
\sigma_4=\left(
             \begin{array}{cc}
               0 & 1 \\
               -1 & 0 \\
             \end{array}
           \right).
\end{eqnarray}
According to our normalization, we find
\begin{equation}
f^{IJ}_{Kr,Ls}=-\textrm{Tr}(Y_{Ls}\left[T^{IJ},Y_{Kr}\right]).
\end{equation}
\indent Generally, the $d=\textrm{dim}(G/H)$ scalar fields $\phi^i$,
$i=1,\ldots , d$ can be described by a coset representative $L$. The
useful formulae for a coset space are
\begin{eqnarray}
L^{-1}t^\mathcal{M}L&=&\frac{1}{2}\mathcal{V}^{\mathcal{M}}_{\phantom{as}IJ}T^{IJ}+\mathcal{V}^\mathcal{M}_{\phantom{as}\alpha}X^\alpha+
\mathcal{V}^\mathcal{M}_{\phantom{as}A}Y^A,\label{cosetFormula}\\
L^{-1} \partial_i L&=& \frac{1}{2}Q^{IJ}_i T^{IJ}+Q^\alpha_i
X^{\alpha}+e^A_i Y^A\label{cosetFormula1}
\end{eqnarray}
where $e^A_i$, $Q^{IJ}_i$ and $Q^\alpha_i$ are the vielbein on the
coset manifold and $SO(N)\times H'$ composite connections,
respectively. $X^\alpha$'s denote the $H'$ generators.
\\
\indent Any gauging can be described by a symmetric and gauge
invariant embedding tensor satisfying the so-called quadratic
constraint
\begin{equation}
\Theta_{\mathcal{PL}}f^{\mathcal{KL}}_{\phantom{asds}\mathcal{(M}}\Theta_{\mathcal{N)K}}=0,\label{theta_quadratic}
\end{equation}
and the projection constraint
\begin{equation}
\mathbb{P}_{R_0}\Theta_{\mc{MN}}=0\, .\label{theta_projection}
\end{equation}
The first condition ensures that the gauge symmetry forms a proper
symmetry algebra while the second condition guarantees the
consistency with supersymmetry.
\\
\indent The T-tensor given by the moment map of the embedding tensor
by scalar matrices $\mc{V}^\mc{M}_{\phantom{as}\mc{A}}$, obtained
from \eqref{cosetFormula}, is defined by
\begin{equation}
T_{\mathcal{AB}}=\mathcal{V}^{\mc{M}}_{\phantom{as}\mc{A}}\Theta_{\mc{MN}}\mathcal{V}^{\mc{N}}_{\phantom{as}\mc{B}}\,
.\label{T_tensor_def}
\end{equation}
Only the components $T^{IJ,KL}$ and $T^{IJ,A}$ are relevant for
computing the scalar potential. With our $SO(8,8)$ generators, we
obtain the following $\mc{V}$ maps
\begin{eqnarray}
\mc{V}_{\mc{A}1}^{ab,IJ}&=&-\frac{1}{2}\textrm{Tr}(L^{-1}J_1^{ab}T^{IJ}),\qquad
\mc{V}_{\mc{B}1}^{ab,IJ}=-\frac{1}{2}\textrm{Tr}(L^{-1}t_1^{ab}T^{IJ}),\nonumber
\\
\mc{V}_{\mc{A}1}^{ab,Kr}&=&\frac{1}{2}\textrm{Tr}(L^{-1}J_1^{ab}Y^{Kr}),\qquad
\mc{V}_{\mc{B}1}^{ab,Kr}=\frac{1}{2}\textrm{Tr}(L^{-1}t_1^{ab}Y^{Kr}),\nonumber
\\
\mc{V}_{\mc{A}2}^{\hat{a}\hat{b},IJ}&=&-\frac{1}{2}\textrm{Tr}(L^{-1}J_2^{\hat{a}\hat{b}}T^{IJ}),\qquad
\mc{V}_{\mc{B}2}^{\hat{a}\hat{b},IJ}=-\frac{1}{2}\textrm{Tr}(L^{-1}t_2^{\hat{a}\hat{b}}T^{IJ}),\nonumber
\\
\mc{V}_{\mc{A}2}^{\hat{a}\hat{b},Kr}&=&\frac{1}{2}\textrm{Tr}(L^{-1}J_2^{\hat{a}\hat{b}}Y^{Kr}),\qquad
\mc{V}_{\mc{B}2}^{\hat{a}\hat{b},Kr}=\frac{1}{2}\textrm{Tr}(L^{-1}t_2^{\hat{a}\hat{b}}Y^{Kr})
\end{eqnarray}
where we have followed the convention of calling the semisimple part
$SO(4)\times SO(4)$ and the nilpotent part $\mathbf{T}^{12}\sim
\mathbf{T}^6\times \mathbf{T}^6$ as $\mc{A}$ and $\mc{B}$ types,
respectively. We then compute the T-tensor components
\begin{eqnarray}
T^{IJ,KL}&=&g_1\left(\mc{V}_{\mc{A}1}^{ab,IJ}\mc{V}_{\mc{B}1}^{cd,KL}+\mc{V}_{\mc{B}1}^{ab,IJ}\mc{V}_{\mc{A}1}^{cd,KL}-
\mc{V}_{\mc{B}1}^{ab,IJ}\mc{V}_{\mc{B}1}^{cd,KL}\right)\epsilon_{abcd}\nonumber \\
&
&+g_2\left(\mc{V}_{\mc{A}2}^{\hat{a}\hat{b},IJ}\mc{V}_{\mc{B}2}^{\hat{c}\hat{d},KL}+\mc{V}_{\mc{B}2}^{\hat{a}\hat{b},IJ}
\mc{V}_{\mc{A}2}^{\hat{c}\hat{d},KL}-
\mc{V}_{\mc{B}2}^{\hat{a}\hat{b},IJ}\mc{V}_{\mc{B}2}^{\hat{c}\hat{d},KL}
\right)\epsilon_{\hat{a}\hat{b}\hat{c}\hat{d}},
\\
T^{IJ,Kr}&=&g_1\left(\mc{V}_{\mc{A}1}^{ab,IJ}\mc{V}_{\mc{B}1}^{cd,Kr}+\mc{V}_{\mc{B}1}^{ab,IJ}\mc{V}_{\mc{A}1}^{cd,Kr}-
\mc{V}_{\mc{B}1}^{ab,IJ}\mc{V}_{\mc{B}1}^{cd,Kr}\right)\epsilon_{abcd}\nonumber \\
&
&+g_2\left.\mc{V}_{\mc{A}2}^{\hat{a}\hat{b},IJ}\mc{V}_{\mc{B}2}^{\hat{c}\hat{d},Kr}+\mc{V}_{\mc{B}2}^{\hat{a}\hat{b},IJ}
\mc{V}_{\mc{A}2}^{\hat{c}\hat{d},Kr}-
\mc{V}_{\mc{B}2}^{\hat{a}\hat{b},IJ}\mc{V}_{\mc{B}2}^{\hat{c}\hat{d},Kr}
\right)\epsilon_{\hat{a}\hat{b}\hat{c}\hat{d}}\, .
\end{eqnarray}
\indent The scalar potential can be computed by using the formula
\begin{equation}
V=-\frac{4}{N}\left(A_1^{IJ}A_1^{IJ}-\frac{1}{2}Ng^{ij}A_{2i}^{IJ}A_{2j}^{IJ}\right)\label{potential}
\end{equation}
in which the metric $g_{ij}$ is related to the vielbein by
$g_{ij}=e^A_{i}e^A_j$. The $A_1$ and $A_2$ tensors appearing in the
gauged Lagrangian as fermionic mass-like terms are given by
\begin{eqnarray}
A_1^{IJ}&=&-\frac{4}{N-2}T^{IM,JM}+\frac{2}{(N-1)(N-2)}\delta^{IJ}T^{MN,MN},\label{A1}\\
A_{2j}^{IJ}&=&\frac{2}{N}T^{IJ}_{\phantom{as}j}+\frac{4}{N(N-2)}f^{M(I
m}_{\phantom{as}j}T^{J)M}_{\phantom{as}m}+\frac{2}{N(N-1)(N-2)}\delta^{IJ}f^{KL\phantom{a}m}_{\phantom{as}j}T^{KL}_{\phantom{as}m}\,
.\nonumber \\
& &\label{A2}
\end{eqnarray}
\indent Finally, we repeat the condition for supersymmetric critical
points. The residual supersymmetry is generated by the eigenvectors
of the $A^{IJ}_1$ tensor with eigenvalues equal to $\pm
\sqrt{\frac{-V_0}{4}}$.
\section{Explicit forms of the scalar potential}\label{potential_form}
For $SO(4)_{\textrm{diag}}$ invariant scalars, the potential is
given by
\begin{eqnarray}
V&=&4 e^{6 a_1} g_1^2 \cosh^2(a_3-a_4) \cosh^2(a_3+a_4) \left[5 \cosh[2 (a_1-2 a_3)]+8 \cosh(4 a_3)\right.\nonumber \\
& &+5 \cosh[2 (a_1+2 a_3)]-4 \cosh(2 a_1) \left(7+2 \cosh(2 a_3) \cosh(2 a_4)\right)+2 \cosh(4 a_4)\times \nonumber \\
& &\left. \left(\cosh a_1 -3 \sinh a_1\right)^2-6 \left(\cosh(4
a_3)-4 \cosh(2 a_3) \cosh(2 a_4)-6\right) \sinh(2 a_1)\right]
\nonumber \\
& &
+4 e^{6 a_2} g_2^2 \cosh^2(a_3-a_4) \cosh^2(a_3+a_4) \left[5 \cosh[2 (a_2-2 a_3)]-8 \cosh(4 a_3)\right.\nonumber \\
& &
+5 \cosh[2 (a_2+2 a_3)]-4 \cosh(2 a_2) \left(7+2 \cosh(2 a_3) \cosh(2 a_4)\right)+2 \cosh(4 a_4) \times \nonumber \\
& &
\left.\left(\sinh a_2 -3 \cosh a_2\right)^2-6 \left(\cosh(4 a_3)-4 \cosh(2 a_3) \cosh(2 a_4)-6\right) \sinh(2 a_2)\right]\nonumber \\
& &-2 e^{a_1+a_2+6 (a_3+a_4)} g_1 g_2 \left[86 \cosh(a_1+a_2)-64 \cosh(a_1-a_2) \cosh(2 a_3)+\cosh(2 a_3)\times \right.\nonumber \\
& & \cosh(6 a_4) \left(\cosh a_1 -3 \sinh a_1 \right) \left(3 \cosh
a_2 -\sinh a_2 \right)+16 \cosh a_1  \cosh(4 a_3) \sinh a_2
\nonumber \\
& & +\cosh(2 a_4) \left[-64 \cosh(a_1-a_2) +\cosh(6 a_3) \left(3
\cosh a_1 -\sinh a_1 \right) \times\right.
\nonumber \\
& &\left.\left(\cosh a_2 -3 \sinh a_2 \right)+2 \cosh(2 a_3) \left(37 \cosh(a_1+a_2)-19 \sinh(a_1+a_2)\right)\right]\nonumber \\
& &-66 \sinh(a_1+a_2)+2 \cosh(4 a_4) \left[8 \cosh a_2  \sinh a_1 +\cosh(4 a_3) \left(\sinh(a_1+a_2)\right.\right.\nonumber \\
& &\left.\left.-3 \cosh(a_1+a_2)\right)\right]+\left[25
\cosh(a_1+a_2)-27 \cosh a_2  \sinh a_1 +2 \cosh(4 a_3)\times
\right.\nonumber
\\
 & &\left.\left(3 \cosh a_1 -\sinh a_1 \right) \left(\cosh a_2 -3 \sinh a_2 \right)-35 \cosh a_1  \sinh a_2 \right] \sinh(2 a_3) \sinh(2 a_4)\nonumber
 \\
 & &+2 \left(\sinh(a_1+a_2)-3 \cosh(a_1+a_2)\right) \sinh(4 a_3) \sinh(4 a_4)+\sinh(2 a_3) \sinh(6 a_4)\times \nonumber \\
 & &\left. \left(3 \cosh a_2 -\sinh a_2 \right)\left(\cosh a_1-3 \sinh a_1 \right)
 \right].
\end{eqnarray}
The potential for $SU(2)^+_L\times SU(2)^-_L$ invariant scalars is
given by, in notation of section \ref{flow},
\begin{eqnarray}
V&=&128\left[g_1^2e^{2b_1}\cosh^2b_2\cosh^2b_3\cosh^2b_4\left(e^{b_1}\cosh
b_2\cosh b_3\cosh b_4-1\right)^2\right.\nonumber
\\
& &
\left.+g_2^2e^{2b_5}\cosh^2b_6\cosh^2b_7\cosh^2b_8\left(e^{b_5}\cosh
b_6\cosh b_7\cosh b_8-1\right)^2\right].
\end{eqnarray}


\end{document}